# Challenges and Practices in Quantum Software Testing and Debugging: Insights from Practitioners

JAKE ZAPPIN, William & Mary, USA
TREVOR STALNAKER, William & Mary, USA
OSCAR CHAPARRO, William & Mary, USA
DENYS POSHYVANYK, William & Mary, USA

Quantum software engineering is an emerging discipline with distinct challenges, particularly in testing and debugging. As quantum computing transitions from theory to implementation, developers face issues not present in classical software development, such as probabilistic execution, limited observability, shallow abstractions, and low awareness of quantum-specific tools. To better understand current practices, we surveyed 26 quantum software developers from academia and industry and conducted follow-up interviews focused on testing, debugging, and recurring challenges. All participants reported engaging in testing, with unit testing (88%), regression testing (54%), and acceptance testing (54%) being the most common. However, only 31% reported using quantum-specific testing tools, relying instead on classical and manual methods. Debugging practices were similarly grounded in classical strategies, such as print statements, circuit visualizations, and simulators, which respondents noted do not scale well. The most frequently cited sources of bugs were classical in nature: library updates (81%), developer errors (69%), and compatibility issues (62%)—often worsened by limited abstraction in existing quantum SDKs. These findings highlight the urgent need for better-aligned testing and debugging tools integrated more seamlessly into the workflows of quantum developers. We present these results in detail and offer actionable recommendations grounded in the real-world needs of practitioners.

## 1 INTRODUCTION

Quantum software engineering (QSE) is gaining prominence as Quantum Computing (QC) transitions from theoretical exploration to real-world application. This shift is driven by the limitations of classical computing hardware [4] and the rise of accessible QC platforms such as IBM's Qiskit [71], Google's Cirq [79], Microsoft's Q# [54], and Xanadu's PennyLane [5]. Developers, physicists, engineers, and software practitioners are now actively writing quantum programs, which, like any software, require robust testing and debugging to ensure correctness and reliability [7, 63].

However, the unique characteristics of QC, such as superposition and entanglement, pose novel engineering challenges [46]. These challenges are further exacerbated by the immaturity of quantum platforms and ecosystems, which lack the comprehensive support and stability found in classical environments [56]. The probabilistic nature of quantum computation, platform instability, frequent

Authors' addresses: Jake Zappin, William & Mary, Williamsburg, VA, USA, azappin@wm.edu; Trevor Stalnaker, William & Mary, Williamsburg, VA, USA, twstalnaker@wm.edu; Oscar Chaparro, William & Mary, Williamsburg, VA, USA, oscarch@wm.edu; Denys Poshyvanyk, William & Mary, Williamsburg, VA, USA, denys@cs.wm.edu.

breaking changes in libraries, and compatibility issues (both software and hardware) present significant obstacles to developers. However, despite these challenges, our understanding of how quantum developers test and debug their applications remains limited [82, 90, 91].

To address this gap, we conducted an empirical study to investigate the current state of QSE, focusing primarily on testing and debugging practices among practitioners. The study involved a survey of 26 quantum software developers from both academia and industry, followed by in-depth interviews with four selected participants. Our research focused on how quantum developers test their applications, the debugging techniques they employ, and the recurring issues they encounter during development.

Our findings reveal several key insights. Although all participants reported participating in testing, most commonly through unit, regression, and acceptance tests, only 31% reported using quantum-specific testing tools. This highlights a gap between academic tool development for quantum applications and their practical adoption in industry, a gap often attributed to limited awareness, weak integration, and inadequate documentation. Developers also reported frequent reliance on manual debugging strategies, such as print statements and circuit visualizations, approaches reminiscent of early classical software engineering practices. Notably, the most significant challenges developers faced were not rooted in quantum-specific phenomena but in familiar classical issues such as developer mistakes (*e.g.*, syntax errors, API misuse, logic errors, *etc.*) platform instability, and integration problems with libraries.

These findings have several important implications for the future of QSE. First, they underscore the need for better integrated, scalable, and user-friendly tools that are specifically tailored to the unique requirements of quantum contexts and developer workflows. Second, they emphasize the importance of improving documentation, community support, and abstraction layers to bridge the persistent divide between academic tool development and industry adoption. Finally, they highlight the potential to significantly enhance quantum developer productivity and software quality by addressing pain points in testing and debugging workflows that are becoming increasingly well understood through emerging empirical research.

In summary, this paper makes the following contributions:

- A novel and comprehensive survey of quantum software developers from both industry and academia, focusing on the challenges they face in testing, debugging, and developing quantum applications;
- Follow-up interviews with four of these developers, providing deeper insights into the testing and debugging strategies they employ and the software engineering obstacles they encounter;
- An analysis of survey responses, offering both quantitative and qualitative insights into current practitioner practices, as well as recurring issues in QSE;
- An examination of key areas where existing tools and methodologies fall short, revealing the need for specialized quantum testing and debugging solutions;
- An investigation and analysis of recurring challenges faced by practitioners, including their underlying causes and how they manifest in quantum applications and algorithms;
- A set of actionable recommendations and insights—based on survey and interview findings—to improve QSE practices, particularly in testing, debugging, and tool adoption;
- A publicly available, anonymized dataset of survey responses to support future research aimed at advancing QSE practices [1].

## 2 BACKGROUND

QC represents a transformative paradigm shift within the fields of CS and, more specifically, software engineering [32]. As a multidisciplinary domain, it leverages the foundational principles of quantum mechanics to process information in fundamentally novel ways that diverge significantly from classical computing approaches [8, 92]. This section provides an overview of QC and its challenges.

### 2.1 The Qubit, Superposition, and Entanglement

At the core of QC lies the *qubit*, the fundamental unit of quantum information. Unlike classical bits, which exist strictly as `0` or `1`, qubits take advantage of the principles of *superposition* and *entanglement* to exist in a combination of states simultaneously [33]. This capability allows quantum computers to represent and process multiple possibilities in parallel, enabling them to perform many calculations at once and offering the potential for exponential speed-ups in certain computational tasks compared to classical systems [72].

In addition to superposition, the phenomenon of entanglement provides another critical advantage. Entanglement creates an interdependence between the states of two or more qubits such that an operation or measurement on one qubit instantaneously influences the state of the other, regardless of distance. This behavior, famously described by Einstein as "spooky action at a distance," exemplifies the non-local correlations that distinguish quantum systems from classical ones [21]. Together, superposition and entanglement enable quantum computers to perform highly complex operations in tandem, unlocking the potential to solve problems that are otherwise infeasible for classical systems [51, 72, 75].

The utilization of superposition and entanglement gives rise to powerful quantum algorithms capable of substantially reducing the time complexity of computationally intensive problems. For example, Shor's algorithm for prime number factorization achieves polynomial-time factoring, a significant improvement over the exponential time required by classical methods [11]. This breakthrough has profound implications for cryptography and security systems, which are highly reliant on the difficulty of factoring large numbers. Similarly, quantum algorithms have shown promise in other domains, such as optimization, quantum chemistry, and machine learning, where classical approaches struggle due to computational limitations [51, 75].

As QC technology progresses, researchers anticipate not only the development of more efficient quantum algorithms that can replace their classical counterparts, but also the ability to tackle problems that remain entirely intractable for classical machines [61]. However, realizing this potential requires overcoming significant challenges in the design, implementation, and testing of quantum algorithms and applications, underscoring the importance of robust QSE practices.

### 2.2 Quantum Computing Limitations and the NISQ-Era

Despite its transformative potential, QC currently faces significant technical challenges, primarily due to the inherent instability of qubits. Qubits are highly susceptible to environmental interference, thermal noise, and other intrinsic factors that cause errors. One of the most significant of these is *decoherence*, the loss of the integrity of a quantum state over time [68]. These challenges necessitate the implementation of error correction techniques, which often rely on error-correcting qubits. However, such approaches introduce additional complexity and computational overhead that can further exacerbate errors [67]. The achievement of true fault tolerance, where errors are effectively suppressed, remains one of the central challenges in QC and is expected to require quantum processors capable of managing millions of qubits [28]. However, this level of scalability remains far out of reach given the limitations of today's quantum hardware. Modern quantum devices, known as Noisy Intermediate-Scale Quantum Systems (NISQ), currently operate with up to a few thousand qubits, as demonstrated by recent advancements from IBM and other companies [15, 35, 87]. In particular, IBM has announced plans to release a 4,000-qubit system by the end of 2025 [36].

The present NISQ era is characterized by quantum computers that contain dozens—to—hundreds of noisy qubits and lack practical error correction mechanisms [10]. Unlike fault-tolerant systems,

NISQ devices rely on probabilistic methods to achieve useful computations. Programs must be run multiple times—known as *shots*—to obtain statistically meaningful results, which adds to the complexity of testing and debugging [78]. Despite these limitations, NISQ systems are already showing practical advantages in running certain quantum algorithms, offering evidence that useful computation is possible even on today's NISQ-era hardware. For example, algorithms such as the Variational Quantum Eigensolver (VQE) have shown how NISQ devices can approximate solutions to complex problems that are otherwise computationally expensive or even intractable for classical systems [10].

As the backbone architecture of current quantum software development, NISQ systems play a dual role: they enable the practical exploration of quantum algorithms while simultaneously exposing the challenges inherent to QC. In fact, reliance on NISQ devices has a direct impact on QSE practices. Developers must design software capable of tolerating noise, mitigating decoherence, and accommodating the probabilistic nature of outputs. These constraints push the boundaries of traditional software development and require the creation of innovative tools, methods, and frameworks specifically tailored for quantum environments. Practical experience with NISQ systems is essential for understanding both the limitations and possibilities of QC, and it continues to inform the development of quantum software designed to scale with future fault-tolerant hardware.

### 2.3 Establishing the Foundations: Quantum Software Engineering

QSE is an emerging discipline dedicated to addressing the distinct and complex challenges involved in developing applications and algorithms for quantum computers. Unlike classical computing, which operates within a deterministic framework, QC is based on principles such as superposition and entanglement, introducing fundamentally different requirements for software design, development, testing, and debugging [56, 63]. As quantum systems continue to evolve and integrate into real-world applications, QSE will become an increasingly essential discipline for ensuring the reliability, maintainability, and scalability of quantum software. By offering systematic and disciplined approaches to quantum software development, QSE aims to bridge the gap between theoretical quantum algorithms and their practical implementation in real-world applications.

One of the foremost challenges in QSE lies in the development of effective debugging and testing methodologies tailored specifically to quantum systems. Unlike classical software, which can be validated through deterministic outcomes, quantum software operates within inherently probabilistic environments. The principles of superposition and entanglement make debugging significantly more complex, as developers cannot rely on classical techniques—such as setting breakpoints or inspecting program state at runtime—to locate errors [46, 82]. These approaches are often inadequate in the quantum context, where observing a quantum state collapses it, offering limited insight into the system's behavior. This necessitates the creation of new tools and methods capable of accommodating the non-deterministic and fragile nature of quantum computations.

In addition to the probabilistic nature of quantum systems, developers must contend with quantum errors, which arise from phenomena such as decoherence, gate fidelity errors, and noise introduced by physical qubits. These issues are particularly prominent in the NISQ-era, where quantum hardware remains noisy and error-prone. Ensuring software correctness in such environments presents substantial challenges for testing and validation. QSE must therefore focus on advancing specialized frameworks and debugging tools that allow developers to efficiently detect, diagnose, and correct these quantum-specific errors. Such frameworks are essential for enhancing confidence in the correctness and robustness of quantum applications as they scale toward increasingly complex problems.

The challenges faced by QSE are further compounded by the relative immaturity of QC platforms and tools. Unlike classical software development, which benefits from mature ecosystems of tools, environments, and best practices, quantum software development lacks standardized frameworks and development workflows [31, 38, 39]. Current quantum platforms remain fragmented, with developers working across tools and libraries such as Qiskit, Cirq, and PennyLane, each with differing levels of support, features and maturity. This fragmentation, combined with the rapid pace of innovation in the quantum space, hinders the widespread adoption of best practices [7, 64]. As the discipline of QSE matures, there is an urgent need to develop and standardize methodologies, frameworks, and tools that can integrate seamlessly across diverse quantum platforms. This will ensure that quantum software is not only robust and scalable, but also adaptable to future advancements in hardware and algorithms.

Establishing these methodologies is critical not only for advancing QSE, but also for supporting the broader integration of QC into domains such as cryptography, chemistry, and optimization. However, to effectively guide the evolution of QSE, it is essential to study the real-world challenges faced by quantum developers—their workflows, testing practices, debugging strategies, and recurring obstacles and challenges they encounter. By grounding QSE in empirical insights from practitioners, the discipline can develop tools, frameworks, and methodologies that are not only theoretically sound, but also practically relevant, adoptable, and adaptable. Addressing challenges such as the difficulty of validating non-deterministic behavior, the absence of runtime state inspection, hardware-induced noise, and the fragmentation of development tools and libraries requires a deep understanding of how developers currently navigate these issues. Through continued research into these practices, QSE can unlock the full potential of quantum technologies and accelerate their successful deployment across diverse real-world applications. This study aims to contribute to that effort by providing empirical insights into the challenges, testing strategies, and debugging practices of real-world quantum software developers.

## 3 RELATED WORK

Research into QSE is gaining momentum as the community works to establish the foundations of a field still in its early stages. In this section, we first review efforts focused on defining the field of QSE and its core challenges. We then examine specific studies that address the practical aspects of quantum software development, particularly testing, debugging, and developer experience.

### 3.1 Foundational Work in Quantum Software Engineering

QSE is an emerging discipline that remains in its infancy, mirroring the early stages of QC itself. As QC continues to evolve, so does the need for structured methodologies in designing, developing, and maintaining quantum software. Unlike classical software engineering, which has undergone decades of refinement, QSE is still defining its foundational principles, tools, and best practices. Although progress has been made, challenges remain in key areas such as testing, debugging, and quality assurance. Early research has reflected this growing need, with Zhao [92] providing a foundational framework for QSE and highlighting the need for methodologies tailored to the unique constraints of QC. Expanding on this, Piattini *et al.* [63] present the Talavera Manifesto, which advocates for structured software development processes, quality assurance, and integration of hybrid quantum-classical systems.

Recognizing the parallels between the early struggles of classical software engineering and the current state of QSE, Moguel *et al.* [56] argues that lessons from classical software crises can inform the evolution of quantum software methodologies. The authors stress the need for high-level programming abstractions and standardized development practices. Similarly, Murillo *et al.* [58] outline a roadmap to advance QSE, emphasizing the importance of adapting classical software engineering methodologies while addressing the unique characteristics of QC. Their work underscores the need for collaboration between QC researchers and software engineering professionals to establish coherent and scalable frameworks, workflows and tooling in QSE.

From a more practical perspective, Haghparast *et al.* [30] examine the challenges faced by quantum software developers, particularly with respect to debugging, testing, and integration of QSE with modern agile methodologies. The authors highlight the difficulties of applying iterative development techniques to quantum software due to hardware limitations and the stochastic nature of quantum execution. Although their work provides a conceptual analysis of the friction between agile practices and quantum development constraints, our study presented in this paper builds on this by offering empirical evidence drawn directly from practitioners.

Together, these works illustrate the nascent yet rapidly advancing landscape of QSE research. Foundational efforts are beginning to define key areas of the field, including how bugs and issues are identified, how testing and debugging are approached, and how quantum developers navigate emerging workflows. Although still in its early stages, QSE is steadily evolving through interdisciplinary contributions that aim to establish practical methodologies and toolchains grounded in both theoretical insight and empirical understanding.

### 3.2 Bug and Issue Classification

Understanding the nature of bugs in quantum software is no doubt crucial to developing effective testing and debugging techniques. As a result, much effort has been devoted to identifying and classifying bugs in quantum systems. In one of the first papers on this topic, Campos *et al.* [14] recognized the importance of standardized resources and advocated for a benchmark dataset of quantum bugs. Aoun *et al.* [46] performed an empirical study on 125 open source quantum software projects on GitHub, finding that quantum software projects are more buggy and costly to fix than classical software projects. The authors identified 13 different types of bugs that occur in 12 quantum components, with program anomaly, configuration, and data type/structure bugs being the most common. Paltenghi *et al.* [59] conducted an empirical study of bugs in QC platforms, identifying quantum-specific bug patterns and emphasizing the need for new quantum-specific techniques to prevent, detect, and fix bugs. Their work analyzed 223 real-world bugs from 18 open-source projects. Furthermore, Zhao *et al.* [94] provided a dataset of 36 bugs collected from the Qiskit quantum computing platform. Expanding on this, Zhao *et al.* [93] also studied bugs in Quantum Machine Learning (QML) frameworks, providing insights into the challenges developers face and a dataset of labeled real-world bugs.

Our previous work, Zappin *et al.* [90], took a different approach by characterizing hybrid quantum–classical issues from a developer's perspective, drawing on discussions from the Xanadu Discussion Forums and Quantum Computing Stack Exchange. In addition to providing a robust taxonomy of bugs and issues encountered by quantum developers, our prior study revealed that a substantial proportion of stability problems in hybrid quantum–classical (HQC) applications stem from programmer errors, while a notable percentage also arise from platform issues, highlighting the need for more robust error-handling mechanisms in QC platforms. Bensoussan *et al.* [9] developed a complementary taxonomy of real-world faults in HQC systems by mining GitHub repositories, identifying recurring fault patterns in classical-quantum interaction code, orchestration logic, and framework-level abstractions, with an analytical emphasis on repository-level artifacts rather than developer-reported experiences.

### 3.3 Testing Quantum Application

Despite QC and QSE being in their early days, researchers have begun to explore various testing methodologies to ensure the reliability of quantum software. Drawing parallels with classical methods, Miranskyy *et al.* [55] suggested adapting well-established testing approaches for classical programs to test quantum programs. However, testing methodologies that are more tailored to the complex nature of QC have been proposed. For example, Long *et al.* [47] presented methods for checking the equivalence, identity, and unitarity of quantum programs, which supported black-box testing. That team also developed a testing process that encompasses unit testing and integration testing for multi-subroutine quantum programs [48]. Wang *et al.* [83] presented QuanFuzz, an approach to automatically fuzzy test quantum programs, achieving greater branch coverage compared to traditional test models. In a different approach, Paltenghi *et al.* [60] proposed MorphQ, a metamorphic testing approach for QC platforms, to detect bugs in Qiskit. Mendiluze *et al.* [52] presented Muskit, a mutation analysis tool for quantum programs developed with Qiskit, to help assess the quality of test cases. Furthermore, Fortunato *et al.* [23] evaluated the effectiveness of input- and output-based coverage criteria for testing quantum programs, while Ali *et al.* [6] extended this work by defining input–output coverage criteria and applying mutation analysis to assess their effectiveness. And, Abreu *et al.* [3] explored metamorphic testing techniques for quantum programs that incorporate oracle functions—black-box components commonly used in quantum algorithms—and defined metamorphic relations specifically tailored to testing such quantum oracles.

### 3.4 Debugging Strategies for Quantum Programs

Debugging quantum programs can require specialized techniques to address the unique challenges posed by quantum mechanics and the rudimentary quantum programming tools currently available. Huang and Martonsi [34] proposed assertions of the quantum program based on statistical tests on classical observations. Complementing this, Li *et al.* [45] suggested a projection-based method for adding assertions to quantum programs at run-time. Sato *et al.* [73] presented a bug-localization method for quantum programs, identifying key program characteristics, such as quantum control flow and measurement patterns, that can guide the efficient detection of faulty code segments. Furthermore, Chen *et al.* [16] implemented a technique called AutoQ for verification and bug localization in quantum circuits, which was able to find injected bugs in various huge-scale circuits. To improve the process of determining quantum states, Witharana *et al.* [88] presented a framework to automatically generate quantum assertions to check different quantum states, assisting developers in debugging quantum applications.

### 3.5 Development Methodologies

As quantum software development matures, structured workflows will become more essential to manage the complexity of quantum applications. Several studies have emphasized the necessity of defining systematic processes in QSE, particularly to ensure seamless integration between quantum and classical components. Weder *et al.* [86] proposed a lifecycle of quantum software development that incorporates quantum-specific phases, such as quantum circuit design and simulation, along with classical stages such as requirement analysis, testing and deployment, underscoring the need for coordinated execution across both paradigms. Similarly, Murillo *et al.* [57] stressed that a well-defined development process is crucial to ensuring maintainability, scalability, and quality in quantum software. Their roadmap outlined key challenges and proposed methodologies for workflow standardization in QSE. Furthermore, Khan *et al.* [39] explored the adaptation of agile workflows for QSE, discussing how iterative approaches can be applied despite the probabilistic nature of quantum computation. Most recently, Upadhyay *et al.* [81] analyzed the evolution and maintenance of QC repositories, revealing a rapidly expanding developer community—with a 200% increase in repositories and a 150% increase in contributors since 2017—while also highlighting a predominance of perfective over corrective commits, suggesting that debugging and maintenance practices may not yet be fully integrated into prevailing quantum development workflows.

### 3.6 Understanding Industry Practices and Challenges in QSE

Despite the growing body of research on QC and QSE in recent years, a significant gap remains between academic efforts and the realities faced by quantum software developers in industry. As illustrated above, much of the existing work in QSE has been theoretical or methodological, focusing on frameworks, taxonomies, and adaptations of classical software engineering principles. Although these contributions are valuable, they do not always align with the practices and challenges encountered by practitioners in real-world quantum application development.

Recent surveys have begun to shed light on the experiences of quantum developers, but their scope has remained broad, often addressing general software engineering topics without deeply engaging with the most critical challenges quantum developers face in application design, testing, debugging, and day-to-day development workflows. In particular, the survey by Stefano *et al.* [19] and the work by Jimenez-Navajas *et al.* [37] provide insight into QSE practices, yet they do not focus their investigations specifically on the recurring issues faced by developers when testing, debugging, and managing HQC workflows. Our study differs in its emphasis: It is directed explicitly at quantum developers and focuses on practical challenges related to identifying, diagnosing, testing, and resolving bugs and other issues in quantum applications.

Our prior work [90, 91] first raised this concern by analyzing online developer discussions and observing that, while academic research often emphasizes high-level methodologies and theoretical constructs, practitioners tend to focus on immediate, hands-on challenges—such as debugging quantum circuits, dealing with the lack of robust testing tools, navigating library compatibility issues, and managing the complexity of hybrid quantum-classical integration. However, those earlier findings were exploratory in nature. To investigate this divergence more systematically, we conducted a comprehensive survey and follow-up interviews with quantum software developers from both academia and industry. By focusing our study on the first-hand experiences of developers, particularly in relation to testing, debugging, and workflow challenges, we aim to help bridge the gap between research and practice. Our findings shed light on real-world difficulties faced during quantum software development, including tool adoption barriers, manual debugging strategies, and limited testing infrastructure. These insights provide a foundation for more developer-informed QSE methodologies and underscore the importance of aligning academic efforts with the evolving needs of quantum practitioners.

## 4 STUDY DESIGN

This study aims to provide a systematic examination of QSE practices, with a particular focus on testing, debugging, and recurring issues and challenges that developers and practitioners face when writing quantum applications. Through a survey and follow-up interviews with quantum developers from industry and academia, we seek to identify common methodologies, assess tool adoption, and uncover recurring challenges that impact quantum software reliability and efficiency. By analyzing how practitioners approach these critical aspects of quantum development, our study offers empirical insights that can inform future advancements in QSE, guiding the creation of better tools and methodologies tailored to the unique demands of QC. To this end, our study answers the following research questions (RQs):

**RQ1:** *What practices and methodologies do developers adopt, and what challenges do they encounter, when testing quantum software?* This question explores the testing strategies employed by quantum software developers—including techniques such as unit testing, regression testing, formal verification, and quantum-specific methods—as well as the tools, both classical and quantum-specific, that support these practices. In addition to identifying common testing methods and tools, this RQ aims to uncover the practical challenges developers face when testing quantum software. By examining how testing is conducted and which tools are used, this RQ seeks to determine which methods are effective, which are underutilized, and how well current practices and tools address the unique challenges posed by quantum systems.

**RQ2:** *What debugging practices and methods do developers employ, and what challenges do they face, when debugging quantum software?* Here, we focus on the debugging approaches used by developers to diagnose and resolve issues in quantum software. This includes examining whether developers rely on traditional techniques—such as manual code inspection, logging, and breakpoint-based inspection—or employ specialized quantum debugging methods and tools. It also examines how developers adapt their debugging approaches to cope with probabilistic execution. Additionally, this RQ examines key challenges developers face when debugging quantum software, including limited tool support and the inability to inspect quantum state at runtime.

**RQ3:** *What bug patterns and issues do developers encounter when developing quantum software?* This question investigates the types of bugs and issues commonly encountered by quantum software developers and explores their root causes. By identifying recurring patterns—such as errors caused by library updates, quantum hardware instability, or developer mistakes—this RQ provides insight into the most pressing challenges, their sources, and how they manifest in practice.

Our study, including the survey questionnaire, the participant identification procedure, and the survey and interview protocols, was approved by our institution's ethics review board. An overview of our methodology can be seen in Figure 1.

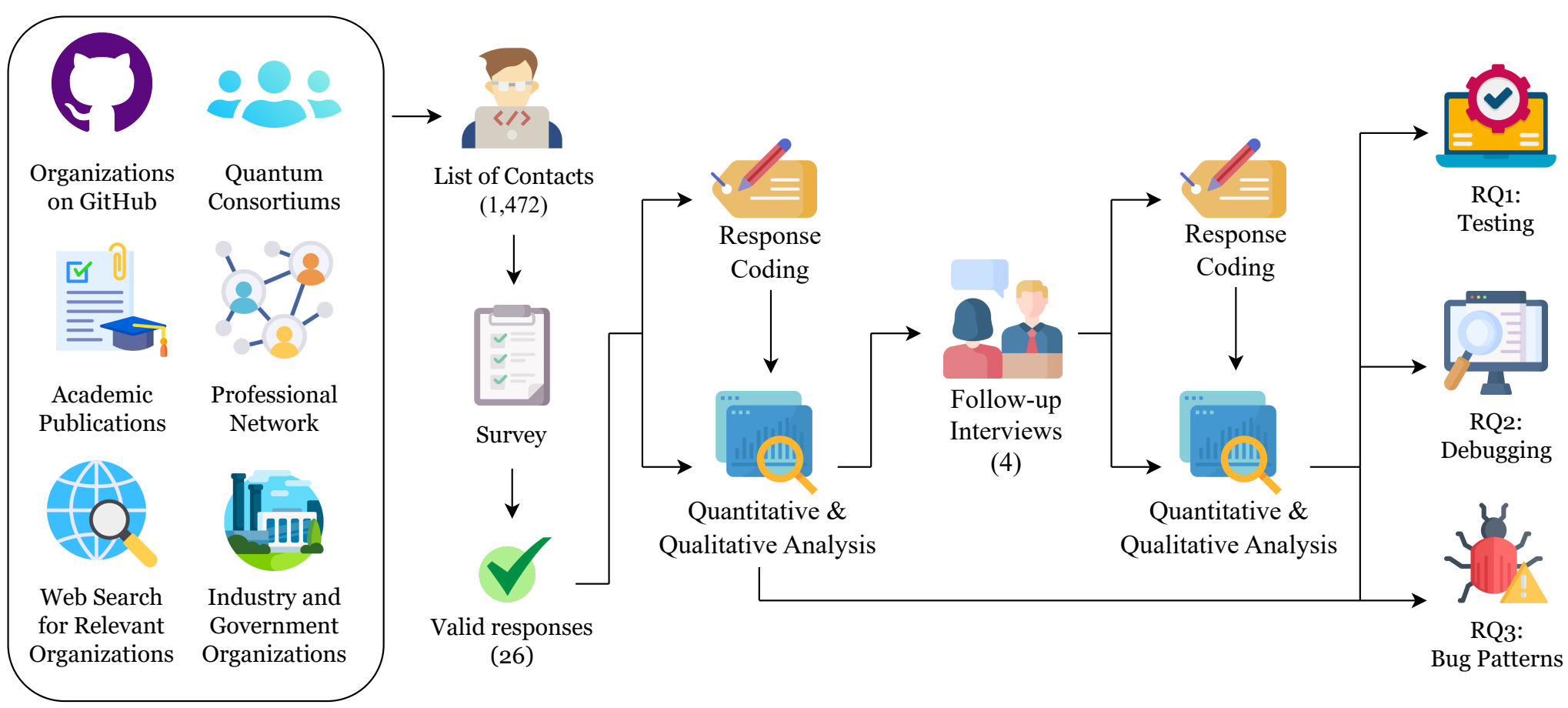

Fig. 1. Overview of Study Methodology

### 4.1 Survey Design

The survey for our study was crafted following general guidelines [29], SE-specific best practices [40–44, 62] and through multiple collaborative sessions by the authors. We used the Qualtrics platform [2] to design, administer, and collect responses for the survey. As shown in Figure 2, the survey consisted of 63 questions, employing a variety of formats to capture diverse types of data. These included multiple choice questions, free-answer questions, Likert scales, and sliding scales. The survey was designed to be completed in approximately 20-30 minutes, balancing the need for thoroughness with the practical consideration of the participants' time. Participation was entirely voluntary, and no compensation was offered to the participants of the survey.

The survey was primarily structured to explore QSE practices in three main categories: (i) testing; (ii) debugging; and (iii) recurring bugs and broader development issues. The first question of the survey ensured that the respondents had prior experience with quantum software development

**Disclaimer, research procedure, participation risks, confidentiality, contact person, and research protocol info**

**Consent form**

**Quantum Software Engineering Experience (1 question)**

(E1) Experience developing, maintaining, or testing quantum software [yes/no]

**Development Environment Background (2 questions)**

(DEB1) Types of quantum computing [multiple choice]
(DEB2) Use of platrforms and libraries [multiple choice]

**Testing (18 questions)**

(T1) General approach to testing quantum software [open-ended]
(T2) Testing methods [multiple choice]
(T3) Types of quantum-specific testing [multiple choice]
(T4) Whether encountered flaky tests [yes/no]
 if yes: (T5) Strategies to mitigate flakey tests [open-ended]
(T6) Whether assertions are used in testing [yes/no]
 if yes: (T7): Types of assertions employed [open-ended]
(T8) Whether formal verification is performed to check correctness [yes/no]
 if yes: (T9): Methods and techniques for formal verification [open-ended]
(T10) Tools used for testing [multiple choice]
(T11) Whether code coverage is measured [yes/no]
 if yes: (T12) How code coverage is measured [open-ended]
(T13) Whether APIs offered by quantum libraries are tested [yes/no]
 if yes: (T14) Which APIs are tested [open-ended]
(T15) Percentage of time dedicated to writing test code vs production code [constant sum]
(T16) Whether issues related to non-determinism were encountered [yes/no]
 if yes: (T17) Experience encountering non-determinism [open-ended]
(T18) Primary challenges faced in testing quantum software [open-ended]

## Bugs and Issues (22 questions)

**Bugs Introduced by Libraries and Platforms (4 questions)**

(L1) When library updates occur [multiple choice]
(L2) Awareness of library updates [5-point Likert]
(L3) How often bugs are experienced after an update [5-point Likert]
(L4) Experience with bugs encountered after library updates [open-ended]

**Bug Manifestations (10 questions)**

(B1) How bugs manifest [matrix table]
 (B2) Explanation of other bug manifestations [open-ended]
(B3) Whether there are common observed patterns contributing to crashes [yes/no]
 if yes: (B4) Description of common patterns leading to crashes [open-ended]
(B5) Whether there are common observed patterns contributing to incorrect output [yes/no]
 if yes: (B6) Description of common patterns leading to incorrect output [open-ended]
(B7) Whether warnings from libraries have been observed [yes/no]
 if yes: (B8) Description of common patterns leading to library warnings [open-ended]
(B9) Whether bugs have manifested in other ways [yes/no]
 if yes: (B10) Description of other bug manifestations [open-ended]

**Bug Causes (4 questions)**

(B11) Causes of bugs [matrix table]
 (B12) Explanation of other bug causes [open-ended]
(B13) How bugs are introduced [matrix table]
 (B14) Other ways bugs are introduced [open-ended]

**Bug Types and Frequencies (4 questions)**

(B15) Types of bugs encountered [multiple choice]
(B16) Frequency of bug occurrences across subsystems [constant sum]
(B17) Frequency of various quantum-specific bugs [matrix table]
(B18) Explanation of other quantum-specific bugs encountered [open-ended]

**Debugging (11 questions)**

**Approach to Debugging (5 questions)**

(DB1) Approach to debugging [open-ended]
(DB2) Whether visualization is used in debugging [yes/no]
if yes: (DB3) How visualization is used [open-ended, optional]
(DB4) Whether tools are used in debugging [yes/no]
if yes: (DB5) Tools used in debugging [open-ended]

**Challenges Encountered (6 questions)**

(DB6) Whether issues related to non-determinism occur during debugging [yes/no]
if yes: (DB7) Description of issues related to non-determinism during debugging [open-ended]
(DB8) Difficulty of resolving bugs [5-point Likert]
(DB9) Challenges that prolong bug resolution times [open-ended]
(DB10) Percentage of time dedicated to debugging [slider]
(DB11) Recurring challenges encountered during debugging [open-ended]

**Demographics (9 questions)**

(D1) Nature of development work [multiple choice]
(D2) Domains worked in [open-ended]
(D3) Whether contributed to open-source [proprietary/open-source/both/other]
if open-source or both: (D4) Projects contributed to [open-ended, optional]
(D5) Years of experience in quantum computing [multiple choice with ranges]
(D6) Years of experience in software development [multiple choice with ranges]
(D7) Highest level of education [multiple choice]
(D8) Formal training in quantum computing [yes/no]
(D9) Country where the organization (or the person fr freelance) is based [free-text]

**Whether the respondent wants to be contacted for an interview, and, if yes, contact info**

Fig. 2. Survey Questions

before proceeding, allowing us to focus on participants with relevant hands-on knowledge. The demographic questions at the end of the survey collected information about the backgrounds, professional roles, and contexts in which participants develop quantum applications.

In the testing section, participants were asked about the types of tests they performed—such as unit, regression, and integration tests—as well as the frequency of testing and their use or awareness of quantum-specific testing tools. These questions were meant to capture both prevailing testing practices and areas where tool support is lacking. The debugging section focused on how quantum developers identify and resolve bugs, including their use of print statements, visualization tools, simulators, and other debugging techniques. In the bugs and issues section, participants were asked to describe the causes and manifestations of bugs and other issues they commonly encounter during quantum application development, such as those arising from developer error, platform instability, breaking library changes, or compatibility problems, to name a few examples. Together, these questions provided a detailed view of the obstacles quantum developers face, as well as the practices and strategies they use to navigate them.

To ensure the precision and clarity of the survey questions, as well as the proper functionality of the Qualtrics platform, we conducted a small pilot study with three students from our lab. The participants were PhD candidates with a background in SE and some familiarity with QC. This pilot study allowed us to validate that the survey adequately covered the intended subject areas, ensured that the questions were well understood, and confirmed that the platform and survey logic operated as expected.

A copy of our final survey is included in our replication package [1], providing full transparency and allowing other researchers to replicate, validate, or build on our study.

## 4.2 Participants

To gather data for our survey, we implemented a comprehensive and systematic approach to identify and engage quantum software developers across both industry and academia. This strategy was designed to ensure that our participant pool consisted of individuals actively involved in quantum software development and/or engineering.

*4.2.1 Industry and Government Organizations.* Our first step was to compile a list of companies and government organizations engaged in QC and quantum software development. This was accomplished through various methods. We conducted general Google and Google News searches to identify relevant companies and performed LinkedIn searches to locate companies involved in quantum software development. Furthermore, we reviewed the member lists of quantum consortiums, searched Google Scholar for publications related to QC and quantum software development, and identified companies and government organizations with GitHub repositories related to quantum software and algorithms. In total, we were able to identify 71 organizations that appeared to be working in QC and/or quantum software development that had more than ten employees.

Once we compiled a comprehensive list of companies and organizations—which included 65 private companies (*e.g.*, IBM, Google, NVIDIA, IonQ, Rigetti, *etc.*) and 6 US government agencies (*e.g.*, NIST, NASA, NSA, *etc.*)—we manually performed LinkedIn searches to identify all employees of each company. From this list, we manually filtered and identified quantum software developers and engineers using the following approach. We began by excluding individuals in non-technical positions, such as those in human resources, sales, legal, and upper management, based on their job titles, job description, and/or experience listed on their LinkedIn or organization webpage. Next, we narrowed down the remaining list to focus on individuals working specifically in quantum software and algorithm development. This involved a thorough review of each individual's LinkedIn profile to determine their experience in quantum software development. We examined job titles, skills listed, self-summaries, and descriptions of their academic and professional projects. Particular attention was paid to job titles such as "Quantum Software Developer," "Quantum Researcher," "Quantum Scientist," and "Quantum Tester," as well as references to quantum programming tools (*e.g.*, Qiskit, Cirq, Pennylane, *etc.*) and to a recent history of working on quantum software. Where necessary, we further attempted to validate each individual's experience in quantum software development by reviewing Google Scholar for relevant publications, GitHub for verified contributions to quantum software projects, and company websites for relevant profiles, blog posts, and press releases.

We applied exclusion criteria to ensure the relevance of our participant pool. Individuals lacking professional experience in quantum software development were excluded from the participant pool, such as those with limited quantum project experience (*e.g.*, those whose experience was primarily small school projects, individuals who had only worked on a single quantum-related project as part of a course or hackathon, or those whose LinkedIn profiles indicated only brief exposure to quantum software development without sustained involvement) or those not currently in a quantum development role for more than a year according to their LinkedIn profile. However, exceptions were made for individuals currently employed as software developers. Additionally, we excluded individuals primarily involved in quantum physics or optics who appeared to have no or limited quantum programming experience.

To ensure a comprehensive list of potential survey participants, we also reviewed additional sources to identify any quantum software developers who might have been missed during the LinkedIn search. We reviewed the GitHub repositories for each organization on our list to identify any contributors to quantum software projects not found through the LinkedIn search. Additionally, we also reviewed the organization's website for employee profiles, blog posts, and press releases that reference quantum software developers who were not included in the initial LinkedIn search.

By taking these steps, we compiled a list of 1,397 quantum developers from 71 private companies and government agencies to invite to participate in our survey.

*4.2.2 Academia.* In addition to reaching out to industry professionals, we also focused on engaging individuals and groups from academia involved in QC and QSE research. This effort was multifaceted, relying on our professional network, snowball sampling (*i.e.*, asking potential participants to share the survey with colleagues) and quantum research consortia. We identified and contacted the directors and program managers of 12 consortiums from major universities known for their QC research programs (*e.g.*, The National Quantum Laboratory (QLab), the Duke Quantum Center, *etc.*). We requested that the survey be distributed to faculty and students involved in QSE and development. In total, we compiled a list of 75 individuals working in academia with whom we shared our survey and request for further distribution. Through this, we ensured that our survey reached a broad and diverse group of potential respondents within the academic community.

### 4.3 Survey response collection and analysis

Survey responses were collected using Qualtrics [2]. The survey was kept open for approximately twelve weeks between May 20, 2024 and August 5, 2024. Excluding our pilot study, we received 38 completed survey responses. We removed 12 responses from the analysis where the respondent reported having no experience developing, maintaining, or testing quantum software. These respondents on the surface appeared to be involved with quantum software development, but were typically researchers focused on quantum theory, engineers primarily focused on quantum hardware, or individuals involved in the policy or business aspects of quantum computing.

In total, our final survey results consisted of 26 unique participants experienced in QSE from both industry and academia, as shown in Figure 3. Although this number may appear modest, it reflects the relatively small and specialized nature of the global quantum software development community. Notably, several individuals (more than 10) we invited explicitly declined to participate due to non-disclosure agreements with their employers, government security restrictions, or other confidentiality reasons, a common constraint in this field. Despite these limitations, the participant pool includes a diverse mix of roles, sectors (namely private industry and academia), and geographic locations, helping to reduce selection bias and provide a well-rounded view of developer experiences across the quantum software development landscape.

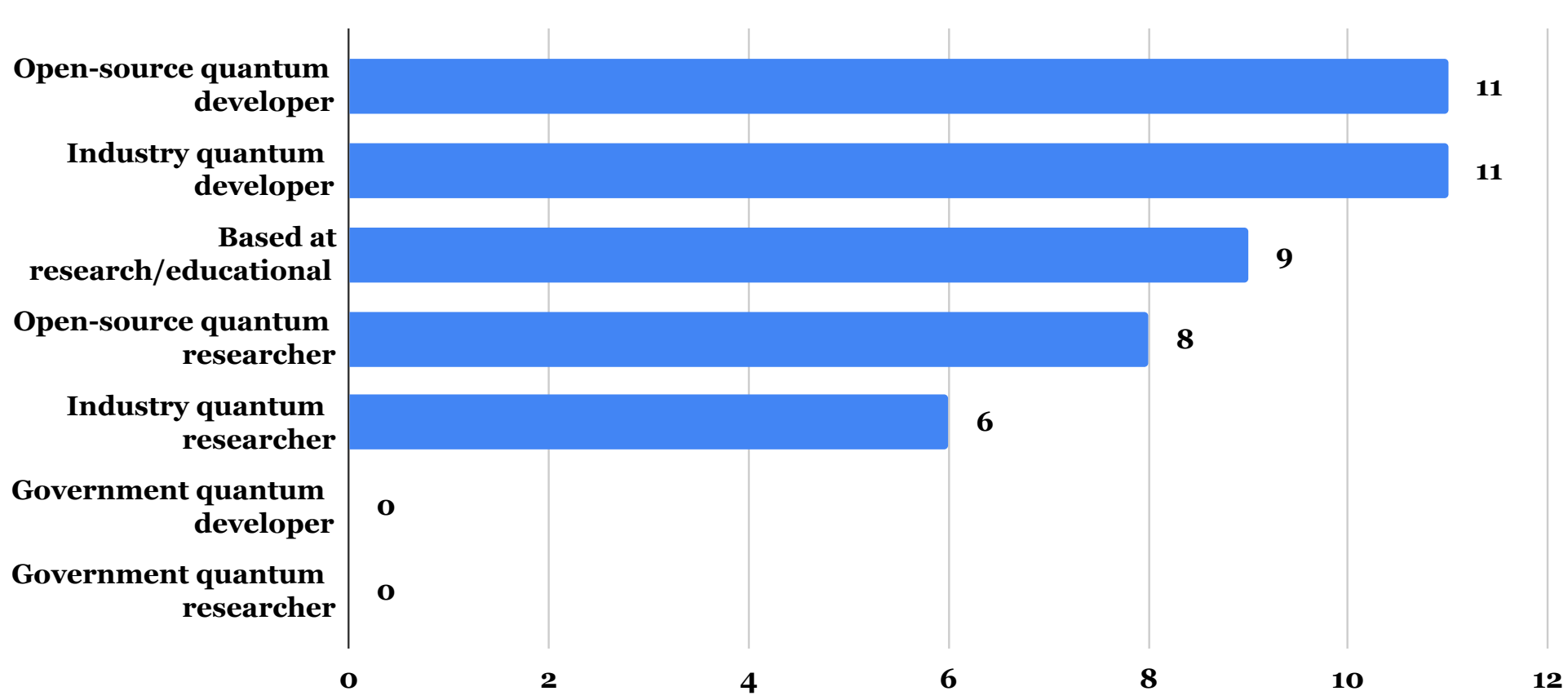

Fig. 3. Background of Respondents (Survey Question D1)

Survey results were compiled using a combination of automated tools and manual analysis to ensure a comprehensive understanding of the data collected. For multiple-choice, Likert scale, and sliding scale questions, the results were automatically generated by Qualtrics. This automated process included the aggregation of responses and the generation of summary statistics, such as mean scores, standard deviations, and response distributions. These automated reports provided an initial overview of the quantitative data, enabling us to quickly identify trends and patterns.

The answers to open-ended questions were analyzed through qualitative open coding [76]:

- *Independent Coding:* Two authors (hereafter "annotators") performed *open-coding* by independently assigning one or more *codes* to each survey response using a shared spreadsheet and codebook. Each annotator independently coded the responses to each of the 11 free-text response questions, adding new codes to the codebook as necessary.
- *Reconciliation Meetings:* Once the initial coding was completed, the annotators met to settle disagreements and consolidate the set of codes. Disagreements were rigorously discussed and resolved. In the event of an irreconcilable disagreement, a third researcher was available to resolve the coding. This step ensured consistency and reliability in the coding process while also mitigating potential bias. We did not base our analysis on inter-rater agreements because multiple codes could be assigned to each response, and no list of codes existed before the start of coding. Our replication package contains our final codebook including definitions [1].

Through this rigorous coding and reconciliation process, the authors systematically analyzed the open-ended responses to extract meaningful insights. The combination of automated quantitative analysis and detailed qualitative analysis ensured that the survey results were both comprehensive and reliable, providing a robust foundation for the study's conclusions.

### 4.4 Participant Demographics

To provide context for our findings, we summarize the demographics of the 26 participants who completed the full survey. As explained above, participants were recruited through targeted outreach to practitioners and researchers in the QSE community.

*Educational Background.* Participants reported a range of academic qualifications. As shown in Figure 4, the majority held advanced degrees, with 18 respondents reporting having a PhD.

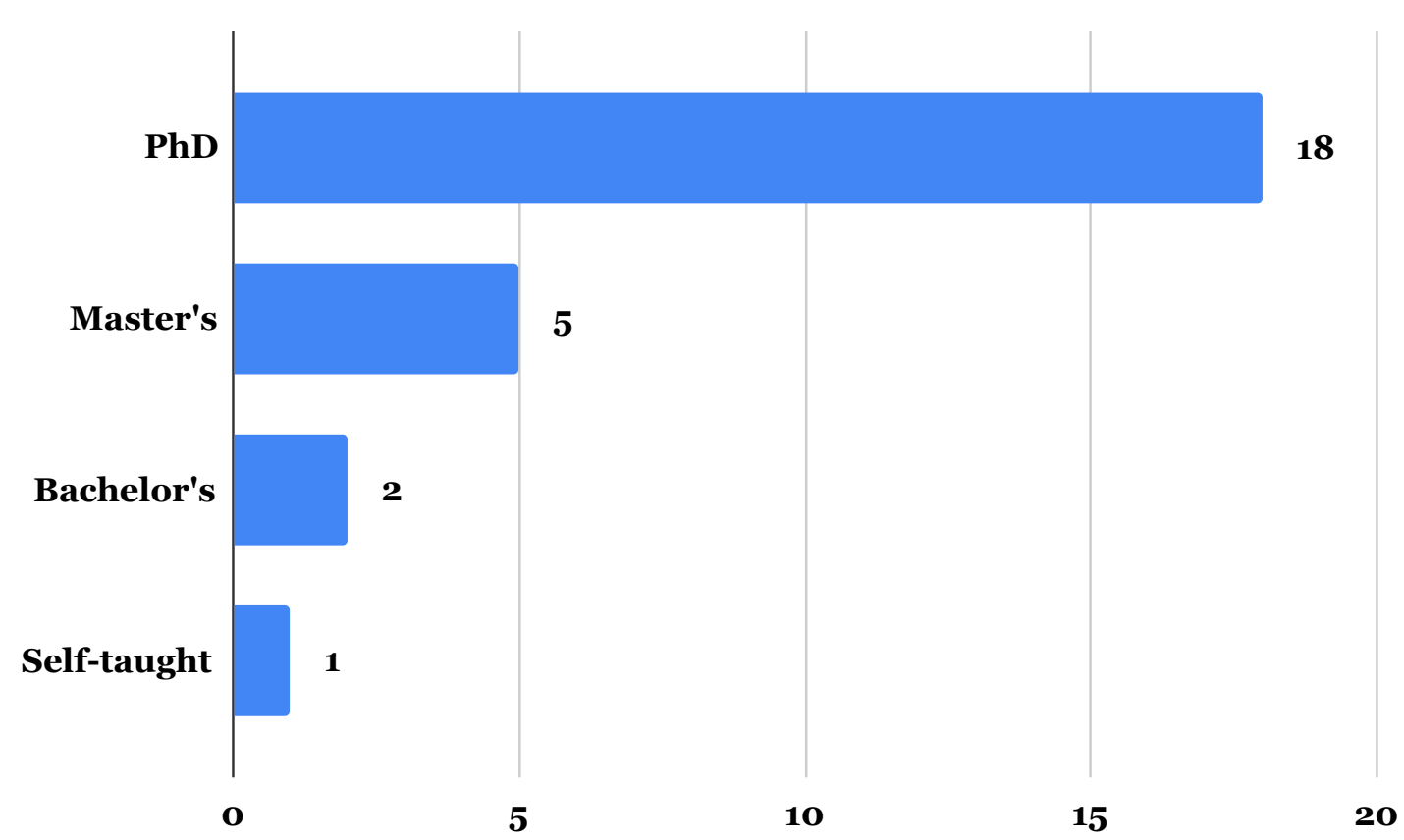

Fig. 4. Respondents' Level of Education (Survey Question D7)

*Experience in the Field.* Respondents also reported varying levels of experience in both QC and software development. Figure 5 shows the distribution of responses based on years of experience in each domain.

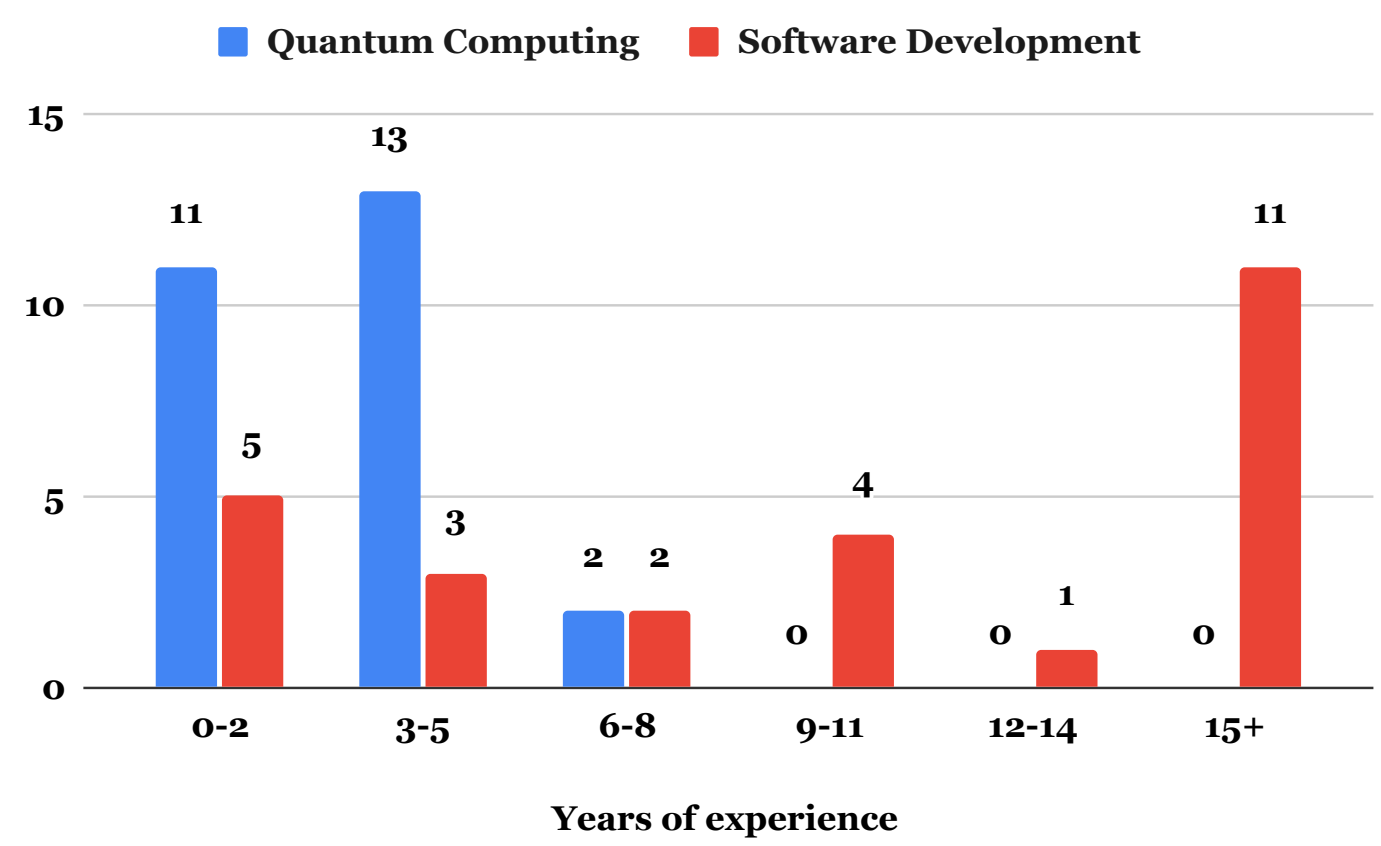

Fig. 5. Respondents' Level of Experience in QC and Software Development (Survey Questions D5 and D6)

*Formal Training*. Finally, we asked participants whether they had received formal training in quantum software development specifically. As shown in Figure 6, 17 out of 26 respondents indicated they had not.

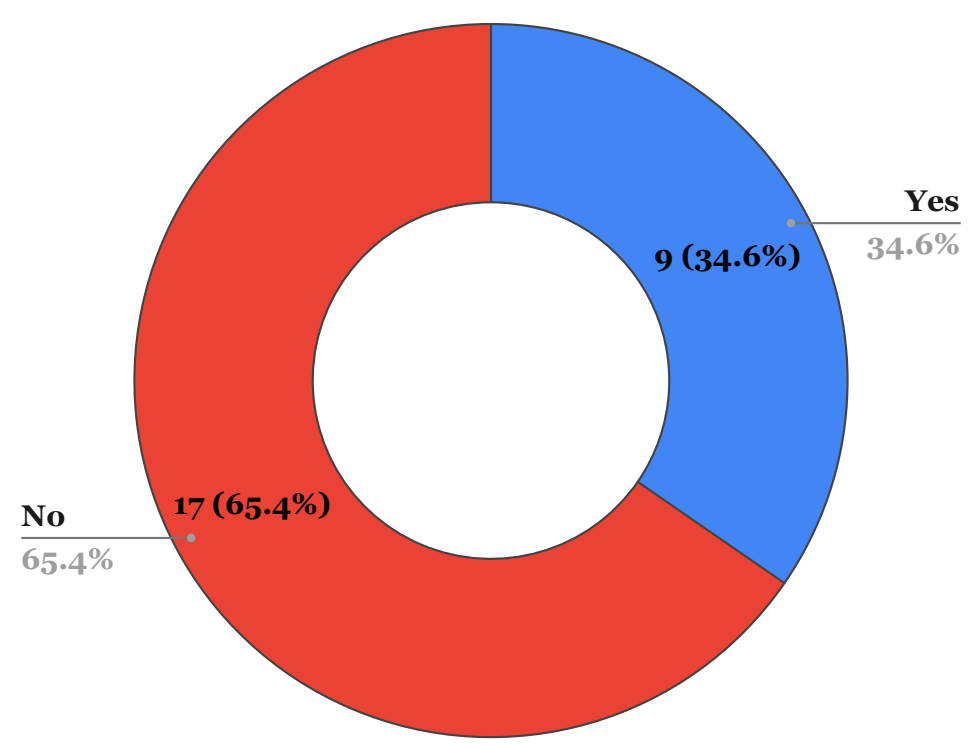

Fig. 6. Respondents' Formal Training in Quantum Software Development (Survey Question D8)

## 4.5 Interviewing Respondents

To complement the survey data and gain deeper insights, two of the authors conducted follow-up interviews with four participants who had indicated a willingness to participate in their survey response. The primary purpose of these interviews was to allow participants to elaborate on their answers, providing additional context about their workflows, testing practices, debugging routines, and challenges they encountered in quantum software development. Additionally, the interviews helped clarify ambiguous responses and fill in gaps identified during the survey analysis, ensuring a richer understanding of the real-world experiences of quantum developers.

The interviews followed a semi-structured format, meaning they were guided by a prepared set of questions, but allowed flexibility for follow-up discussions. Each interview was conducted by one or two authors who engaged the participant in an open-ended dialogue. Prior to each interview, the interviewers reviewed the participant's survey responses and identified specific areas where further elaboration was needed. A personalized list of questions was prepared for each respondent, focusing on topics such as their debugging strategies, experiences with quantum software tools, challenges related quantum development, and their perspectives on the future of QSE.

While these prepared questions provided a foundation for discussion, the interviews were conversational and adaptive rather than rigidly structured. If a participant provided unexpected insights or raised new topics, the interviewers adjusted their approach, asking follow-up questions and exploring related issues in more depth. This flexible format allowed participants to share nuanced perspectives that might not have been captured in the structured survey, leading to a more comprehensive understanding of key challenges they faced and the practices that they employed developing quantum applications.

The interviews were conducted remotely over a two-week period, with each session lasting between 20 to 30 minutes. To ensure accuracy and allow for thorough analysis, all interviews were recorded and transcribed over Zoom or Google Meet. We then conducted a qualitative analysis using open coding to identify recurring themes and patterns across participant responses. These codes were iteratively grouped into higher-level categories aligned with our research questions. The insights gathered from these interviews are reflected throughout Sections 5 to 7, where relevant quotes and remarks are included to support and clarify key findings.

| **Interview Respondent** | **Background** |
|---|---|
| Interview Respondent 1 | A quantum software developer working for a private startup company in industry. The respondent indicated having extensive experience in quantum software development and in particular working with quantum libraries and platforms. |
| Interview Respondent 2 | A professor in academia who self-described as a quantum developer, indicated teaching quantum computing classes at the graduate level, and had written quantum applications in support of those classes. |
| Interview Respondent 3 | An academic quantum developer primarily engaged in quantum computing research. The respondent indicated an extensive experience in writing quantum applications and, in particular, testing tools for quantum applications and algorithms. |
| Interview Respondent 4 | A quantum software developer working in private industry at a startup entity, who indicated having significant experience in writing quantum applications and had most recently been working on quantum machine learning algorithms and applications. |

Table 1. Background of Interview Respondents.

Table 1 summarizes the backgrounds of the four individuals we interviewed. Throughout the paper, we refer to these participants using the notation Interview Respondent 1 through Interview Respondent 4 to preserve anonymity while maintaining clarity.

## 5 RQ1: TESTING METHODOLOGIES USED BY QUANTUM DEVELOPERS

To understand how quantum developers approach testing, we surveyed and conducted follow-up interviews with practitioners about the testing methods and tools they use during quantum software development. We focused on both general testing strategies and the adoption of quantum-specific testing tools, aiming to uncover patterns in current practices and tool usage. We also examined the challenges developers encounter when applying these testing approaches in practice.

## 5.1 Testing Techniques

All 26 survey respondents reported that they engage in some form of testing during the quantum software development process. The survey results indicate a strong reliance on traditional or classical software testing methodologies, with limited adoption of research-driven testing methods and tools proposed in the QSE literature. Developers primarily rely on unit, regression, acceptance, integration, and end-to-end tests to ensure software quality and correctness.

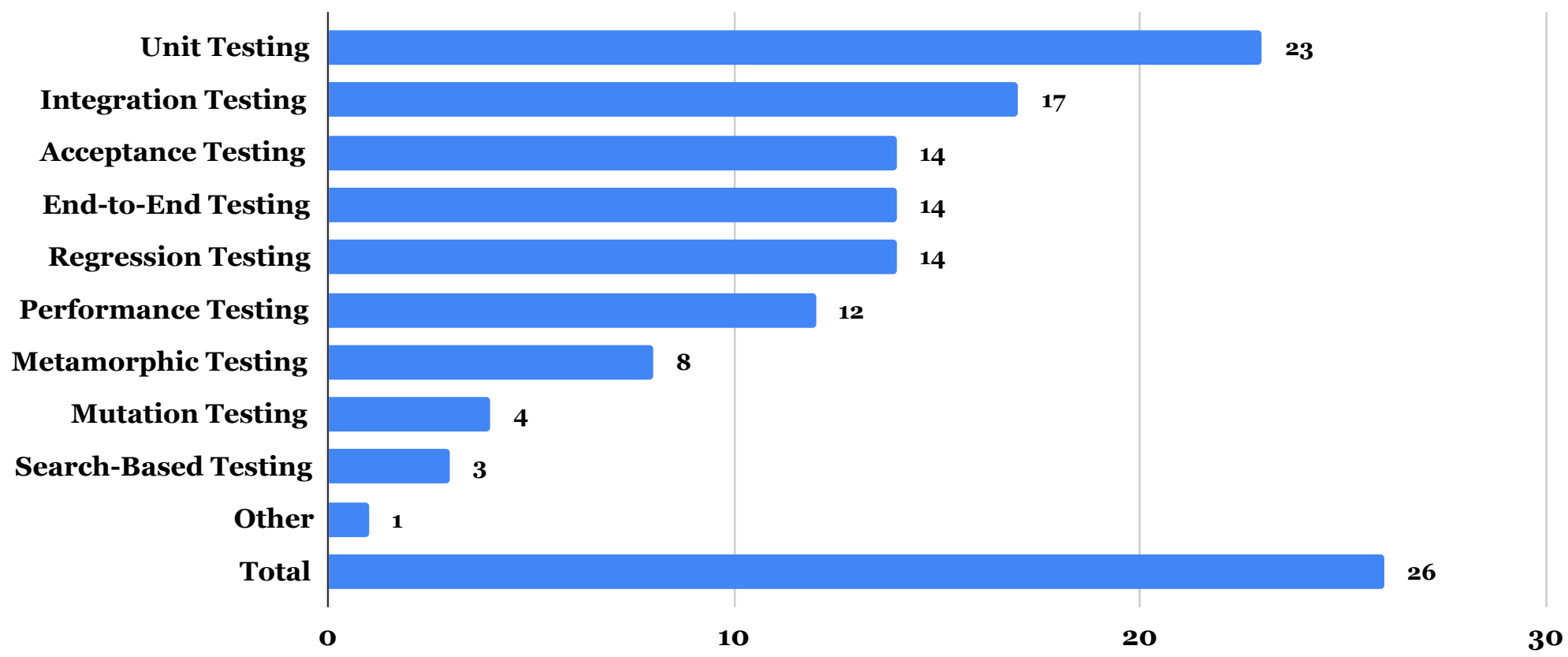

Fig. 7. Types of Testing Used by Survey Respondents (Survey Question T2)

As shown in Figure 7, 23 out of 26 respondents reported using *unit testing*, making it the most commonly used testing methodology. Unit testing is typically used to verify the accuracy of individual functions or modules of an application in isolation. In quantum applications, unit testing is used primarily to validate individual quantum circuits and their components. As Interview Respondent 3 explained: "First, we want to independently check our SPAM—our State Preparation And Measurement—because you cannot really do a circuit unless you know what your starting point is."

In addition to unit testing, respondents reported using a range of other traditional or classical testing approaches for their quantum applications. *Integration testing*, selected by 17 participants, focuses on verifying that different software components work together as intended. This is especially important in quantum applications where multiple circuits are combined and in hybrid systems where classical and quantum subsystems must interoperate correctly. *Regression testing*, used by 14 of 26 respondents, is commonly used to ensure that code updates do not introduce unintended side effects or break existing functionality in the application. In quantum software, regression tests can help validate changes to circuits, but are particularly important when updating quantum libraries or modifying back-end configurations, where subtle shifts in library or API behavior may occur and result in breaking changes to the application. *Acceptance testing* and *end-to-end testing* were also reported by 14 respondents. Acceptance testing typically verifies that the application behaves as expected from a high-level functional perspective—for example, ensuring a new feature produces correct or meaningful output under defined conditions. End-to-end testing validates the complete workflow of a quantum application or algorithm. In hybrid systems, such testing may be especially valuable for verifying the correct interaction between classical control logic, quantum execution backends, and post-processing steps, where errors may only surface when components are exercised together. Lastly, *performance testing*, reported by 12 participants, is used to evaluate aspects such as runtime efficiency, execution latency, and back-end responsiveness—which are particularly critical when relying on cloud-based quantum hardware or simulators executed on classical machines.

Respondents emphasized that the importance of testing quantum applications is to maintain system stability throughout the development lifecycle. As Interview Respondent 4 noted, "Getting those tests set up early can save you a lot of pain later on." Another participant explained, "When we're developing something that's truly new, the standard tools aren't necessarily there ... but automated testing is probably [used] once things are working or near working, and you just want to make sure it hasn't broken."

In contrast, testing methods that have received increasing attention in the quantum software research literature—such as *metamorphic testing* [60], *mutation testing* [24, 52], and *search-based testing* [84]—were far less commonly used in practice based on the survey responses. *Metamorphic testing* checks program correctness by exploiting expected relations between multiple inputs and outputs (*e.g.*, if an input is transformed in a predictable way, the output should transform accordingly). *Mutation testing* introduces small changes, or "mutants," into the code or circuits to evaluate whether existing tests can detect them, thereby measuring the strength of a test suite. *Search-based testing* applies optimization or heuristic search techniques to automatically generate test inputs that maximize coverage or expose faults. Only eight of 26 respondents reported using metamorphic testing, four indicated using mutation testing, and just three reported the use of search-based testing. These results highlight a disconnect between research-driven testing approaches proposed in the quantum software engineering literature and the traditional testing techniques currently employed by practitioners.

Taken together, these findings reveal a divide between traditional and non-traditional testing techniques in practice. The use of traditional testing methods—such as unit, integration, regression, end-to-end, performance, and acceptance testing—was nearly ubiquitous among respondents, reflecting the direct transfer of well-established classical practices into the quantum domain. By contrast, non-traditional approaches that have received recent attention in the quantum research literature—namely metamorphic, mutation, and search-based testing—were rarely reported by respondents. The limited adoption of these methods highlights not only their relative novelty, but also a lack of mature tool support and limited practitioner familiarity, which often led respondents to fall back on *ad hoc* testing, as discussed in Section 5.3. In these instances, developers typically engaged in manual, black-box testing by repeatedly executing circuits with varied inputs and informally judging whether the outputs were plausible. This reliance on exploratory, improvised practices underscores the gap between research-driven proposals and day-to-day development realities, and suggests that testing in quantum software remains largely anchored in traditional techniques while gradually, and unevenly, incorporating more experimental alternatives.

**Finding 1: Reliance on Traditional Testing —** While 100% of respondents engage in some form of testing, 88% use unit testing, 65% use integration testing, and 54% use regression and acceptance testing, underscoring the central role of traditional or classical testing methods in quantum development.

### 5.2 Testing Tools

Despite the widespread use of traditional or classical testing techniques, the survey showed a relatively low adoption of specialized quantum testing tools. As shown in Figure 8, 16 out of 26 respondents indicated that they use no testing tools at all when testing their quantum programs. Of the remaining ten respondents who reported using testing tools, only eight (31%) indicated using formal testing frameworks or libraries designed specifically for quantum software. For example, Qiskit Test Utilities [70], a testing library packaged with the Qiskit platform, was used by only four respondents, despite all respondents reporting use of Qiskit for development. One respondent (Interview Respondent 3) indicated that they employed internal or proprietary testing tools and were unaware of academic quantum testing tools such as Muskit [52], QuCAT [26], and QmutPy [24]. The respondent stated: "I'm not familiar with the academic quantum testing tools [...] One possible answer is ignorance, and another answer would be, we have our own internal ones." Interview Respondent 3 went on to say: "I'll probably look at some of these [other academic testing tools] and maybe if I think some of them are useful, propose [using] them." Taken together, these findings suggest a gap between research-driven innovations and the practical adoption of quantum-specific testing tools, likely due to barriers such as limited awareness, lack of exposure, or minimal integration into widely used quantum development frameworks.

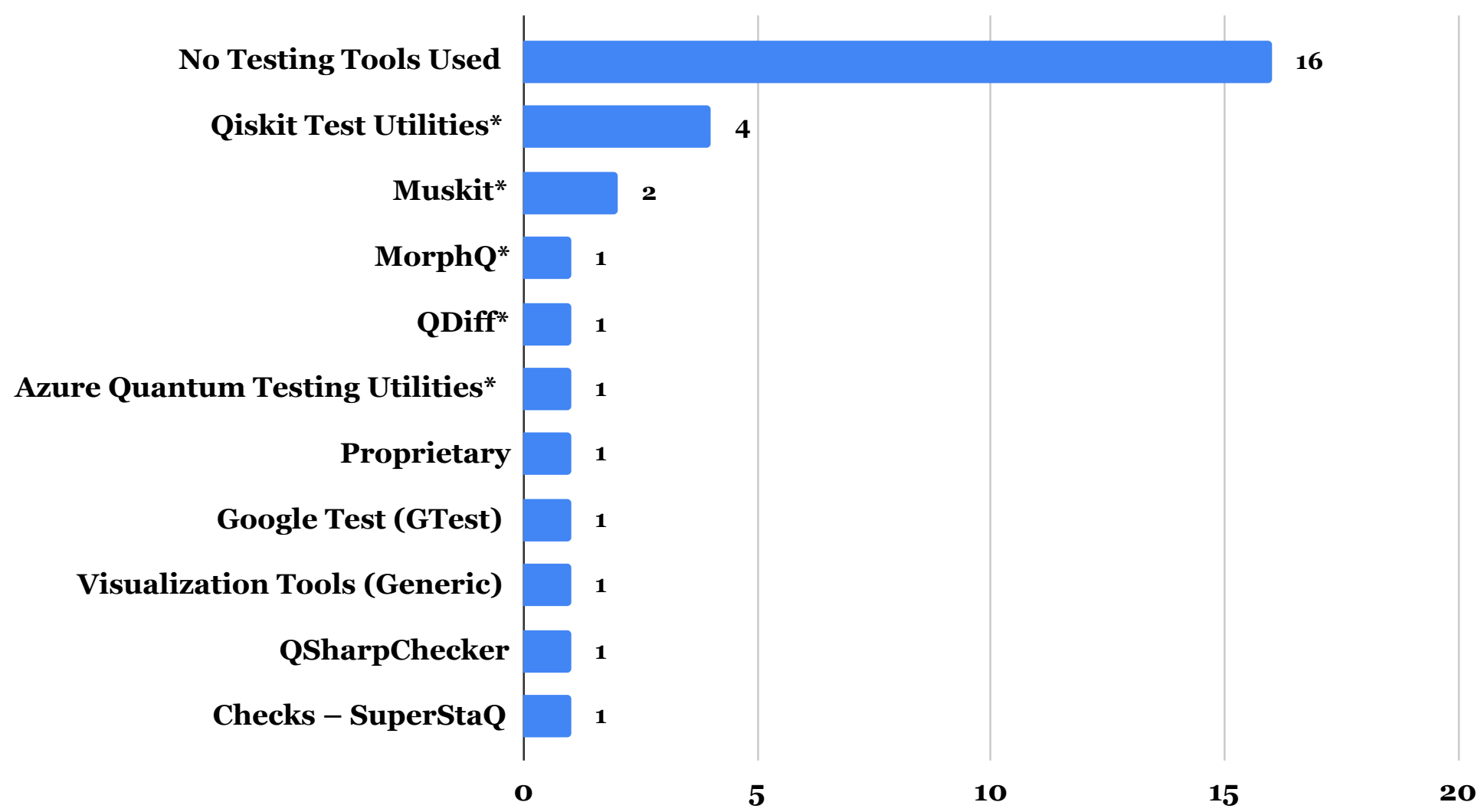

Fig. 8. Testing Tools Used by Survey Respondents (Survey Question T10)
*Testing tool was listed as a multi-select option

**Finding 2: Limited Adoption of Quantum-Specific Tools —** Only 31% of respondents reported using testing tools designed for quantum software, such as formal test utilities, with many citing a lack of awareness of academic and publicly available solutions.

### 5.3 Prevalence of Manual Testing

Manual testing methods were the most predominant way quantum developers tested their applications according to the survey. Of the 16 out of 26 respondents who reported using no testing tools (see Figure 8), nearly all (15 out of 16) indicated using print statements and visual inspection of circuits as part of their testing workflows. This prevalence likely stems from a combination of limited tool support for quantum testing and developers' reliance on familiar, low-overhead techniques adapted from classical software development.

Visualization was specifically mentioned as a helpful tool in low-abstraction contexts, particularly through the use of quantum circuit diagrams and state visualizations (*e.g.*, Bloch spheres or state vector plots), although some participants noted their limitations when working with larger or more complex systems. Interview Respondent 3 remarked, "A lot of the testing is through visualization... I frequently hear from [other] quantum engineers, 'I want to look at it!' And I'm like, no, that doesn't scale! You can't just look at things!" As circuit complexity increases, with more qubits and deeper layers of gates, visualization becomes difficult or impossible, making it a clear limitation in the testing of more advanced quantum applications and algorithms. This highlights the need for more scalable and abstract visualization techniques, similar to graph-based approaches used in classical software engineering to manage complexity by hiding low-level details and emphasizing structural patterns [17, 20, 22].

### 5.4 Quantum-Specific Testing

In addition to the traditional software testing practices reported by respondents and summarized in Figure 7, almost all survey respondents (24 of 26) reported using at least one form of testing specific to the QC context. The technique most commonly used was *circuit testing* [53], selected by 19 of 26 participants, which involves verifying the overall structure and behavior of a quantum circuit by checking that sequences of gates, state preparations, and transformations produce the expected outcomes. *Fidelity testing* [89], used by 17 respondents, measures how closely the actual output state matches the intended or ideal state, often expressed as a probability or overlap value, and is particularly important in noisy environments. Moreover, half of the respondents reported using gate testing [53], which focuses on validating the correctness of individual quantum gates in isolation, ensuring they perform the expected transformations before being integrated into larger quantum circuits.

Beyond these core methods, several respondents indicated the use of other quantum-specific techniques tailored to address noise and entanglement. *Noise and error testing* [53], reported by 10 participants, involves deliberately evaluating quantum programs under noisy or imperfect conditions to understand how hardware or environmental factors affect correctness and stability. *Entanglement testing* [25], selected by nine respondents, is used to verify that qubits become and remain entangled as expected, often through specialized protocols such as the controlled-SWAP test. *Quantum error correction and mitigation testing* [12], reported by 8 participants, refers to techniques that validate the effectiveness of error-correcting codes or noise-mitigation strategies designed to preserve state fidelity on NISQ devices and simulators. Finally, only one participant reported engaging in *communication testing* [27], which focuses on validating the correctness and reliability of transmitting quantum states across communication channels or between quantum devices. Interestingly, two respondents indicated that they performed no quantum-specific testing of their applications at all, instead seemingly relying exclusively on classical testing techniques.

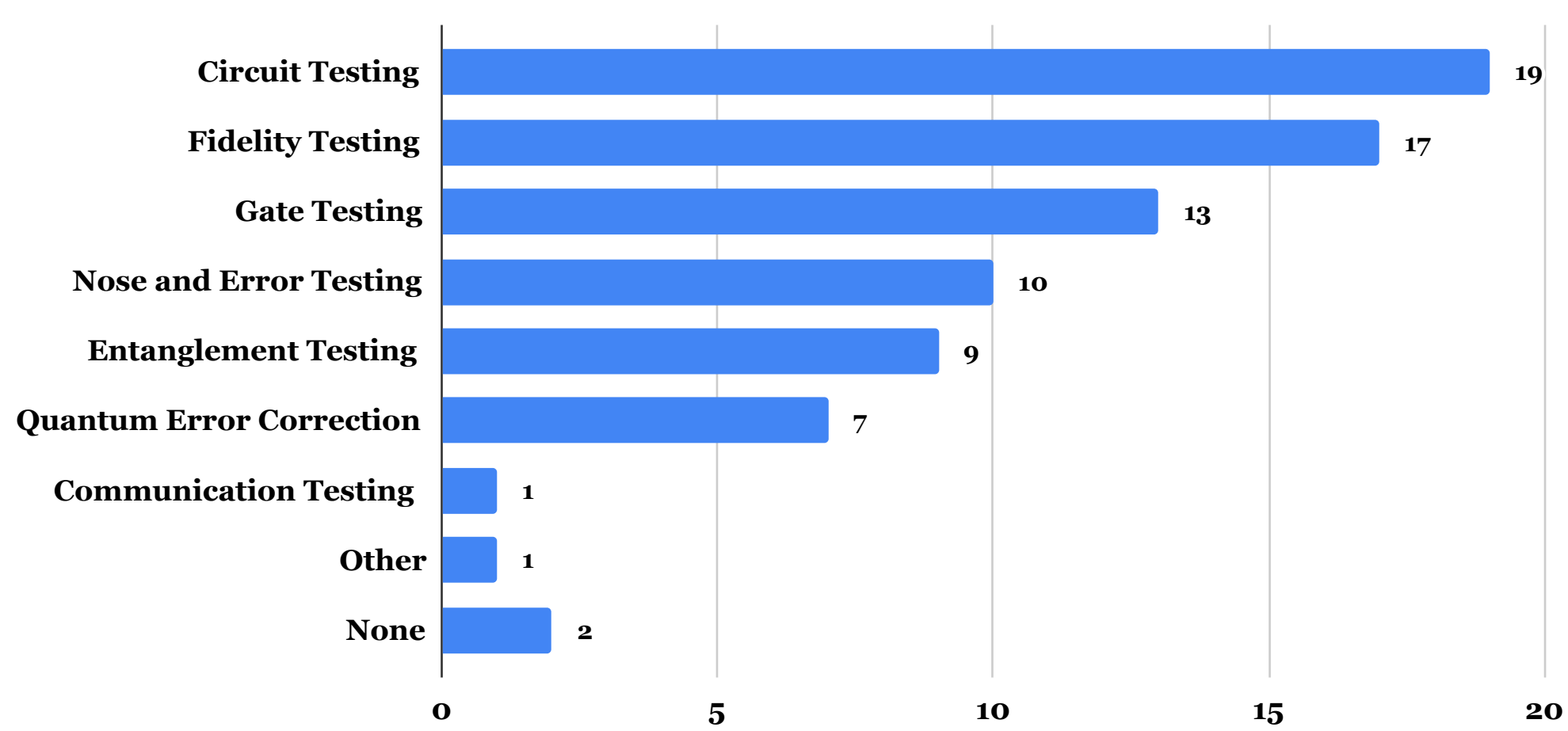

Fig. 9. Types of Quantum-Specific Testing Used by Survey Respondents (Survey Question T3)

These results suggest that in addition to relying on classical testing methods such as unit and regression tests, quantum developers are further burdened with performing complex, domain-specific testing—such as circuit, fidelity, entanglement, and noise testing—that is uniquely required by current NISQ-era devices. Based on the survey responses, it appears that these efforts are often carried out without the support of dedicated tools, leaving quantum developers to improvise strategies for testing challenges and requirements that are fundamentally different from those in classical software engineering. This disconnect between quantum-domain–specific testing requirements and available tooling highlights an urgent need for purpose-built testing frameworks that account for the physical constraints, probabilistic behaviors, and domain-specific characteristics of quantum software.

### 5.5 Assertions, Formal Verification, Coverage, and Flaky Tests

In addition to standard testing strategies, well over half of the respondents reported using programmatic assertions to validate quantum software behavior. Specifically, 16 out of 26 participants indicated that they employ some form of assertions in their test code for quantum applications (see Figure 10). Open-ended responses described these as checks on expected output distributions or validations of intermediate circuit states, seemingly performed under simulation to avoid the state collapse that occurs during actual quantum measurement. Although lightweight and practical, survey responses indicated that these assertions were largely *ad hoc*: crafted manually by developers on a case-by-case basis without relying on standardized assertion libraries or formal specification methods. This suggests a lack of systematic support for assertion-based testing in current quantum development environments.

Figure 10 further shows that formal verification of quantum code was reported by nine out of 26 respondents, suggesting that a subset of developers applies more rigorous testing practices to their quantum applications. Free-text responses described a range of verification approaches, with the most commonly mentioned being comparison against simulation results (four mentions), where developers validated program behavior against that of a trusted simulator. We view simulator-based comparison as a form of verification because developers were not merely observing test outcomes, but explicitly assessing whether program behavior matched a trusted reference. In this context, the simulator serves as a reference model or oracle rather than a general-purpose test harness. Other approaches included analyzing probability distributions, plotting fidelity, applying differentiation techniques, and benchmarking resource usage, each mentioned once. One respondent indicated that their verification approach was “confidential,” suggesting the use of proprietary workflows. Overall, these responses indicate that developers often rely on statistical or simulation-based heuristics in place of rigorous, tool-supported formal verification techniques.

A few respondents also mentioned coverage testing as an important consideration. While our survey did not ask for detailed distinctions, one participant clarified the dual meaning of coverage in quantum applications: “Not sure if this is referring to quantum circuits or the code which produces those circuits. We use lots of conventional tools to measure code coverage. For circuit coverage, we vary the input state to sweep across the solution space of a circuit.” This comment highlights that practitioners may apply both conventional code coverage techniques to the classical components of quantum programs, and circuit coverage approaches that attempt to explore a range of input states in order to assess behavior across the solution space.

In addition, eight out of 26 respondents indicated that they used coverage testing tools to assess test completeness. In classical software engineering, coverage testing is used to measure the extent to which source code has been exercised by a given test suite, providing a quantitative proxy for confidence in test scope. Although high coverage does not guarantee correctness, it helps ensure that major code paths and logic branches are tested, thereby reducing the likelihood of undetected defects. While less common than other forms of testing, coverage metrics provide a quantitative measure of how thoroughly different parts of a quantum program are exercised during testing and may serve as a proxy for confidence in test scope, particularly in simulated environments where most quantum applications are evaluated.

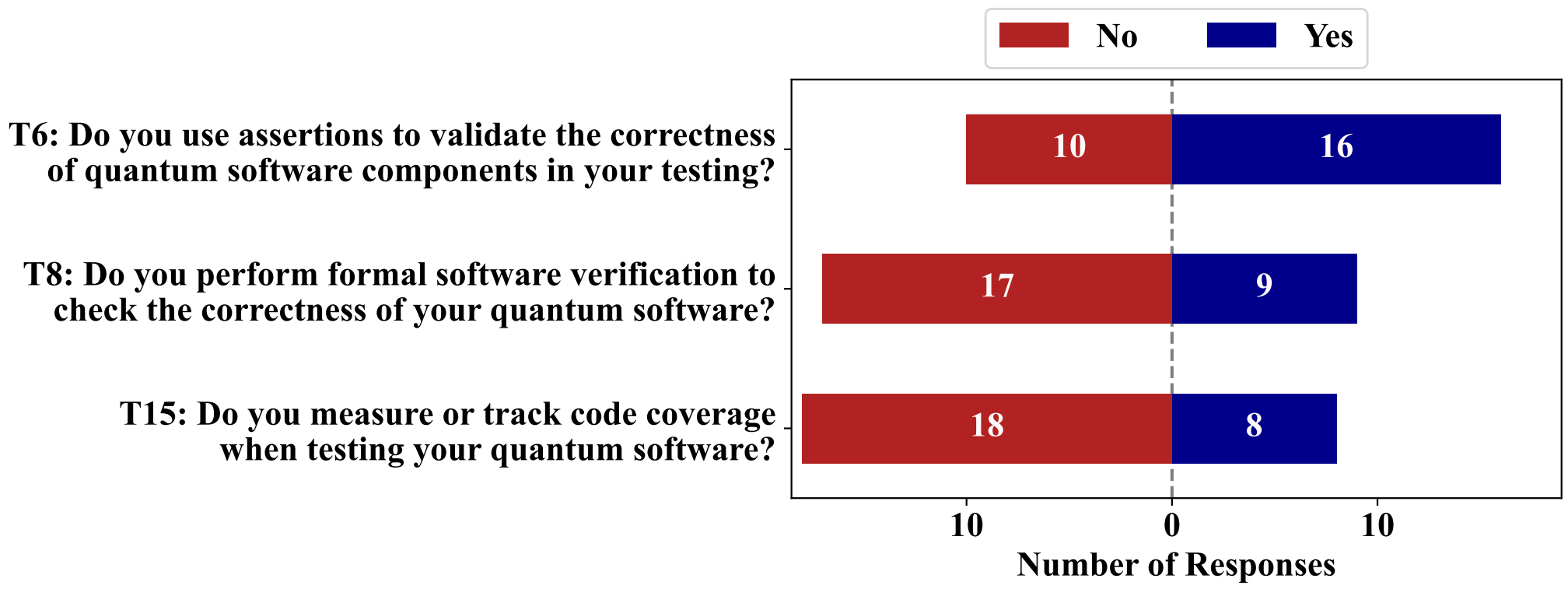

Fig. 10. Testing Methods Employed by Respondents

Furthermore, when asked about flaky tests—tests that fail intermittently without changes to the codebase—14 out of 26 respondents reported encountering such issues. Noise was the most frequently cited cause (four mentions), followed by hardware-related inconsistencies (one mention) and poorly designed unit tests (one mention). To mitigate these problems, respondents reported commonly resorting to strategies such as rerunning tests to gain statistical confidence or rewriting tests to be more tolerant of variation.

Taken together, these findings indicate that quantum developers are actively adapting their testing practices to cope with the uncertainty and instability of current quantum hardware and platforms. The widespread reliance on *ad hoc* assertions and improvised workarounds underscores the absence of mature, well-integrated testing tools tailored to the unique demands of quantum systems. This disconnect highlights a pressing need for dedicated testing frameworks that support more systematic, reliable, and scalable quantum software engineering practices as the field continues to mature.

**Finding 3: Quantum Testing Is Burdensome and Lacks Standardization —** Developers reported performing substantial domain-specific testing—such as circuit testing (73%), fidelity testing (65%), and entanglement or noise validation (35–38%)—in addition to classical testing practices—often without dedicated tool support. Nearly two thirds (62%) rely on manually crafted *ad hoc* assertions, and 54% encounter flaky tests that are typically rerun to gain statistical confidence. Manual visualization remains one of the most commonly used testing methods, but breaks down as circuit complexity increases.

## 5.6 Testing Challenges

In addition to tool limitations, several respondents described challenges related to the inherently probabilistic nature of quantum computation. Non-determinism during testing was flagged by 11 out of 26 participants as a difficulty. Open-text responses revealed a range of related issues, including inconsistencies between runs (one mention), differences between machines (one mention), and result variability over time (one mention). These conditions make it difficult to determine whether observed behavior reflects a bug or expected quantum variability.

The most frequently raised issue—cited by three participants in free-text responses—was that some algorithms require multiple runs to produce statistically meaningful results. This is expected behavior on NISQ-era hardware, where quantum algorithms typically rely on repeated executions, or *shots*, to estimate output probabilities due to inherent noise and measurement variability. Other participants noted inconsistencies such as obtaining different results across runs (two mentions), as well as variations in output between different days or across different machines (one mention each). Additional challenges included testing random measurements (one mention), fluctuations in parameter fidelity (one mention), and difficulty getting circuits to behave as specified (one mention). Three respondents acknowledged the inherent non-determinism of quantum circuits and identified noise as a major contributing factor. Only one participant explicitly reported encountering no issues related to non-determinism.

Beyond non-determinism, respondents reported a broad range of practical challenges. Errors and noise were the most frequently cited issue (four mentions), followed by lack of quantum hardware, slow runtimes, and hardware constraints (two mentions each). Other challenges raised included limited access to complex real-world programs, difficulties with quantum logic, and problems scaling tests to larger systems. Additional difficulties included poor documentation, challenges in writing tests, and achieving meaningful code coverage (one mention each).

Together, these findings highlight the complexity of testing in the NISQ era of QC, where developers must contend with quantum variability, machine-dependent behavior, and noisy execution environments—all in the absence of robust testing frameworks to mitigate or isolate such effects. Testing quantum software is not only technically demanding but also constrained by limited resources, requiring developers to navigate a landscape shaped by hardware limitations, infrastructure gaps, and the inherent unpredictability of quantum systems. To address these challenges, there is a pressing need for testing frameworks, methodologies, and toolchains specifically designed for quantum software—ones that account for probabilistic behavior, support reproducibility, and integrate seamlessly into the current quantum development platforms such as Qiskit and Cirq.

**Finding 4: Testing Quantum Software Remains Challenging —** Developers face non-determinism (42%), hardware noise (15%), and variability across machines and execution time (12%), making it difficult to distinguish bugs from expected behavior. Practical barriers such as slow runtimes, limited hardware access, and a lack of robust testing infrastructure further compound the complexity of testing quantum software in the NISQ-era of QC.

## 5.7 Analysis of Testing Practices Based on Background and Experience

To explore how background and institutional context shape testing practices, we examined responses by quantum experience, software development experience, and organizational affiliation (academia vs. industry). While the sample sizes are modest, clear patterns emerge that highlight differences in the breadth and emphasis of testing strategies.

*5.7.1 Quantum Experience.* Respondents with 0–2 years of quantum experience universally reported the use of unit testing (11 out of 11), often as their primary or sole structured method. Those with 3–5 years of quantum experience (12 out of 13) also reported unit testing but more frequently described employing a broader mix of strategies, including integration testing (nine out of 13), regression testing (eight out of 13), and performance testing (seven out of 13). The small number of respondents with 6–8 years of experience (one out of two reporting unit testing) show similar tendencies, though the sample is too limited for firm conclusions. These findings suggest that as quantum developers gain experience, their testing repertoire expands beyond component-level checks toward system-level validation.

*5.7.2 Software Development Experience.* A parallel pattern appears when considering classical software experience. Respondents with 0–2 years of software experience exclusively reported unit testing (five out of five), with limited mention of broader methods (*e.g.*, three out of five noting end-to-end testing). By contrast, those with 15+ years of experience reported unit testing alongside a much wider set of approaches (ten out of 11), including regression testing (eight out of 11), integration testing (seven out of 11), and performance testing (five out of 11). Intermediate experience groups show similar trends, with nearly all respondents reporting unit testing while increasingly supplementing it with additional testing methods. Taken together, these results suggest that prior exposure to established software engineering practices strongly shapes testing behaviors in the quantum domain.

*5.7.3 Academia vs. Industry.* Comparing respondents based in research or educational institutions (Academia, $n = 9$) with those in commercial or organizational settings (Industry, $n = 17$) reveals a further divide. Both groups reported widespread reliance on unit testing (eight out of nine in academia; sixteen out of 17 in industry), underscoring its role as a baseline practice across organizational contexts. Beyond this, however, industry participants more frequently employed a broader and tool-intensive testing repertoire:

- End-to-End Testing: one out of nine in academia vs. thirteen out of 17 in industry
- Integration Testing: four out of nine in academia vs. thirteen out of 17 in industry
- Performance Testing: none in academia vs. twelve out of 17 in industry
- Regression Testing: three out of nine in academia vs. eleven out of 17 in industry

On average, industry respondents reported 4.88 distinct testing methods per person, compared with 2.67 methods among academics. This suggests that while academics typically confine their practices to a narrower range—often centered on unit or regression testing—industry practitioners integrate multiple testing layers, reflecting the demands of production-level quantum software development and hybrid system integration.

In sum, across experience and institutional subgroups, unit testing is nearly universal. The differences emerge in the breadth of practices: developers with more years of quantum or software experience, and those working in industry settings, consistently employ a richer set of testing methods. These patterns highlight how both individual background and organizational context influence the testing strategies adopted in quantum software development.

### 5.8 Summary

Overall, the results indicate that quantum developers consistently apply traditional testing methods, but must also shoulder the added burden of quantum-specific testing—an effort that is both pronounced and essential in the NISQ-era of QC. Despite the growing importance of these practices, tooling support remains limited, with few developers using formal quantum testing tools or frameworks. As a result, testing in QSE continues to rely heavily on manual processes, internal tools, and improvised solutions. These challenges are further compounded by the inherent non-determinism of quantum systems, machine-specific variability, and the difficulty of distinguishing genuine bugs from expected quantum fluctuations—issues that current tools and methods do little to address. These findings underscore the urgent need for research and development efforts focused on creating testing frameworks and methodologies that are purpose-built for the realities of QSE and the challenges practitioners face. Table 2 presents the key findings of the survey with respect to RQ1.

## 6 RQ2: QUANTUM DEVELOPERS DEBUGGING PRACTICES AND CHALLENGES

We asked survey respondents to describe the strategies and tools they use for identifying and resolving bugs in quantum software. Our analysis reveals a diverse set of debugging practices, a strong reliance on classical debugging techniques, and only limited use of tools designed specifically for quantum systems—mirroring trends observed in testing methodologies. We also examine the challenges developers face when debugging quantum programs.

## 6.1 Prevalence of Manual Debugging

In line with the responses we received regarding testing, Figure 11 shows that a majority of respondents—18 out of 26—reported that they do not use dedicated debugging tools, whether quantum-specific or general-purpose. Instead, many rely on conventional or traditional debugging methods, such as print statements, manual code inspection, and visual circuit representations. As Interview Respondent 4 put it: "At the end of the day, it's usually just as easy to stick in the print statements that you need." Although these approaches may be familiar and accessible, they often fall short in addressing the fundamental complexities of quantum programs, such as probabilistic behavior, state collapse, and limited observability. These strategies tend to be more effective in simulated environments, where quantum states can be inspected deterministically to an extent, but are often impractical or infeasible on real hardware due to noise, measurement constraints, and scalability limitations.

To better understand current debugging practices among quantum developers, we analyzed open-ended responses that describe how they approach debugging in quantum software development (Survey Question DB1). The responses revealed a diverse but largely conventional set of strategies. Many participants described relying on foundational techniques such as inserting print statements into their code(two mentions), inspecting code manually (referenced in 11 responses), or running the same program with different inputs to isolate inconsistencies (five mentions). Others emphasized efforts to reproduce or reduce bugs, including returning to a previously working version of the code (three mentions), simplifying the circuit to a minimal example (three mentions), or isolating the smallest segment that caused the failure (two mentions). This incremental approach—building up and validating the system piece by piece—was reflected in several responses as a practical way to manage complexity and diagnose bugs and other faults in quantum software.

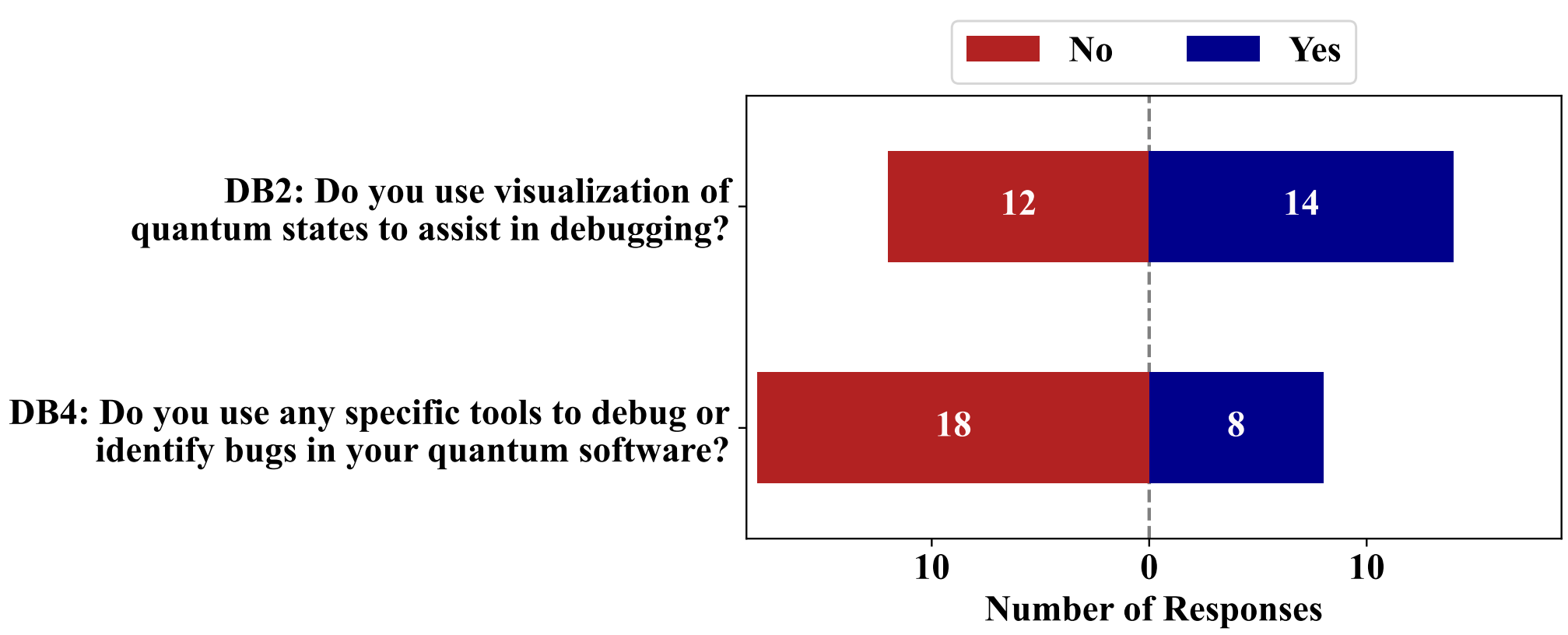

Fig. 11. Use of Debugging Tools by Respondents

Some respondents reported debugging by comparing circuit outputs against mathematical expectations (two mentions), validating results against known templates (one mention), or plotting circuit behavior to better understand state evolution (one mention). It should be noted that these approaches, while precise, often require substantial domain expertise and can be time consuming to execute, particularly when validating complex behaviors across different circuit configurations or theoretical expectations. Six respondents also mentioned that they relied on simulators to help test and debug their applications. These environments allow developers to run quantum circuits without hardware-induced noise or decoherence, making it easier to identify logical errors in isolation. Although quantum simulators still reflect the inherent probabilistic nature of quantum mechanics, they provide a controlled setting where repeated results are consistent and reproducible, features that are especially valuable during early-stage development and debugging. However, a respondent noted that simulators still fall short when trying to model real-time control flow or dynamic behaviors, particularly as applications scale. Peer review (one mention), version control (one mention), and IDE-based debugging tools (two mentions) were also referenced, reflecting influences from classical software engineering workflows. A small number of respondents indicated that they were still learning how to debug quantum software effectively (one mention) or that they were not currently engaged in debugging tasks (three mentions).

These results suggest that quantum developers tend to debug through iterative exploration, simplification, and trial-and-error, rather than relying on dedicated debugging infrastructure or tools—whether general-purpose or designed specifically for quantum software. Debugging was more-or-less described as a process of elimination, seemingly guided by intuition and familiarity with both classical and quantum behavior. Two respondents specifically noted that a first step in their debugging process was ruling out whether a bug had originated in the classical subsystem before investigating quantum-specific causes. Additionally, two other respondents described comparing circuits to known references or validating behavior mathematically in lieu of using formal verification tools. While some of these strategies may be effective in smaller programs or simulation environments, they can become labor-intensive and error-prone as systems grow in complexity. Moreover, techniques such as validating against known templates or mathematical expectations often depend on access to trusted reference implementations and a deep understanding of both quantum mechanics and the underlying hardware—an unrealistic expectation for many future quantum developers. This stands in stark contrast to classical software development, where most practitioners are not expected to possess knowledge of the physical principles underlying the computing systems they use.

**Finding 5: Diverse but Largely Manual Debugging Strategies —** Eighteen of 26 respondents (69%) reported that they do not use dedicated debugging tools. Instead, respondents relied heavily on *ad hoc* classical debugging strategies such as print statements, manual code inspection, visualization, and testing with varied inputs. Only a small number reported employing quantum-specific debugging methods, with many emphasizing process-of-elimination and exploratory approaches rather than formal, tool-supported techniques.

We also asked respondents to identify the tools they use to support debugging (Survey Questions DB4 and DB5). Of those who answered, most cited general-purpose or classical tools, including:

- **Valgrind [74]** and **GDB [77]:** Standard tools for memory and runtime debugging in classical environments.
- **Memory and Time Profilers:** Used to track performance and optimize resource usage.
- **IDE Debuggers:** Such as those built into Visual Studio Code.
- **Plotting Libraries:** Like `matplotlib` [80] and `plotly` [65] used to visualize state evolution.
- **Custom User Interfaces:** Developed internally to interact with quantum systems.
- **Checks-SuperstaQ [13]:** Specialized runtime validation tool for quantum programs.
- **Delta Debugging [66]:** A formal debugging technique cited by one participant, likely referencing a research-based implementation.

Each tool or method listed above received only a single mention by respondents, indicating that debugging practices are highly individualized and that no single tool has gained widespread traction.

Notably, only one of these—Checks-SuperstaQ [13]—is a reusable tool specifically designed for quantum debugging and optimization. The rest are either classical tools, general-purpose utilities, or custom-built solutions. Further, some of these tools, such as Valgrind and GDB, are likely used exclusively for debugging the classical host code in hybrid applications. Others, however—such as plotting libraries and IDE debuggers—could be applied to debugging both classical and quantum portions of programs, for example by visualizing simulation outputs or stepping through the orchestration logic that generates and executes circuits. This limited use of specialized tooling mirrors trends seen in quantum testing and highlights the gap between the needs of quantum developers and the availability and capabilities of the current quantum debugging ecosystem. Our own review of the literature found that only a handful of quantum-specific debugging tools have been proposed to date—including Bugs4Q [94], Qbugs [14], and others discussed in Section 3—highlighting both the limited availability of dedicated debugging tools in QSE and a broader disconnect between academic tool development and industry adoption.

**Finding 6: Sparse Use of Quantum-Specific Debugging Tools —** Despite a wide range of debugging strategies, only one respondent explicitly reported using a tool designed specifically for quantum debugging. Most respondents instead relied on classical tools, custom or proprietary utilities, or general-purpose development environments.

### 6.2 Use of Visualization Tools

As previously mentioned, visualization tools can be used by quantum developers to support reasoning about and inspecting quantum circuit structure as part of the debugging process. Figure 11 shows that 14 out of 26 respondents indicated that they incorporate circuit visualization into their workflows. The most frequently cited tool was Qiskit's `QuantumCircuit.draw()` function (two mentions), though respondents also referenced using Bloch spheres [49] (two mentions), Husimi distributions [18] (one mention), custom Python scripts with SVG output (one mention), and internal visualization platforms (two mentions). Respondents indicated that these visualizations were used for verifying gate sequences (one mention), inspecting output states (two mentions), and diagnosing dependency structures across pipelined components (one mention).

Despite their utility, some respondents pointed to limitations with visualization—particularly when dealing with more complex circuits and quantum programs. As circuits increase in depth or involve more qubits, visual representations can become unwieldy, reducing their effectiveness for tasks that go beyond basic inspection or illustrative purposes. As Interview Respondent 4 explained: "You're thinking at a much lower level … and visualizations help but can't cover the entire complexity." These observations suggest that while visualization is a useful entry-level debugging aid, it lacks the scalability needed to support systematic debugging of complex quantum programs.

**Finding 7: Visualization Tools Are Common but Limited —** Fourteen out of 26 respondents (54%) reported using visualization tools such as `QuantumCircuit.draw()` and statevector or Bloch sphere plots to understand circuit structure or behavior, but noted that these tools do not scale well to larger or more complex quantum systems.

### 6.3 Challenges Faced in Debugging Quantum Software

While respondents reported a range of debugging strategies, many also noted challenges that complicate or prolong the process. Debugging quantum software remains inherently difficult for several reasons—some arising from the probabilistic and opaque nature of quantum computation, and others stemming from infrastructure and ecosystem-level limitations such as hardware access, execution latency, and tooling support.

Only six out of 26 respondents reported encountering issues related to the non-deterministic nature of quantum computation when debugging quantum software (Survey Question DB6). Among those who did encounter issues related to non-determinism, respondents described the unpredictability of output across repeated executions of the same code (two mentions), difficulty identifying the source of failures due to quantum noise (one mention), and test flakiness caused by brittle test designs (one mention) as challenges they have faced. One participant noted, "Rerunning the test will fix it," reflecting the transient and sometimes hardware-dependent nature of these issues. Another explained, "This is rare, but poorly written tests will be brittle to non-deterministic outcomes. The fix is to write tests that are insensitive to this," which highlights how inadequate test design can later complicate debugging efforts by producing inconsistent or misleading results. The respondent further noted that these issues may be even more pronounced when debugging in hardware-based environments (as opposed to quantum simulators), where noise and variability are more difficult to control.

When asked to rate the difficulty debugging bugs in quantum software (Survey Question DB8), nine out of 26 respondents (35%) described the process as either *difficult* or *very difficult*, while the majority (17 respondents, 61%) reported a *neutral* level of difficulty. Only two participants found debugging to be *easy*, and none selected *very easy*. These responses appear somewhat at odds with other findings in our study, which highlighted a range of technical and ecosystem-level challenges that can complicate debugging. One possible explanation is that developers have come to expect a certain level of complexity when working with immature quantum systems and probabilistic behavior, leading them to normalize difficulty as part of the process. In this light, the "neutral" responses may not indicate ease, but rather reflect a resigned or pragmatic outlook on the inherent challenges of debugging quantum software.

Responses to Survey Question DB9 offered insight into specific factors that prolong bug resolution. The most commonly cited challenges were long run times and limited access to quantum hardware (four mentions each), which create significant bottlenecks for testing and debugging, especially when programs must be executed multiple times to validate results. Library-related issues (four mentions) were another major source of difficulty, including dependency conflicts, compatibility problems, and undocumented behavior after updates. Several participants also cited the lack of fault localization tools, frequent changes to interfaces (*e.g.*, APIs), and a proliferation of simulator choices as adding to the complexity of debugging their quantum applications.

Communication and documentation were additional pain points based on the survey responses. Two respondents mentioned poor documentation of quantum libraries or platforms as a hurdle. Two others described the difficulty of explaining or understanding certain types of bugs—challenges that can delay collaboration and problem resolution. Six respondents cited unique obstacles, each mentioned only once, including difficulties drawing circuit diagrams, interpreting noisy input behavior, dealing with changing interfaces, navigating a proliferation of simulator choices, and lacking fault localization support.

These findings underscore that debugging quantum software is a multifaceted challenge. While non-determinism does not appear to be a widespread obstacle for most quantum developers—likely due to the fact that most quantum development work is done on simulators—resource limitations, immature libraries, and weak tooling appeared to be major impediments based on the responses we received. As quantum software scales in complexity, addressing these pain points will be essential to improving productivity and reliability.

**Finding 8: Bug Resolution Is Often Hampered by Ecosystem Limitations —** Developers cited long execution times, hardware constraints, and external library or dependency issues as major barriers to resolving bugs. Communication gaps and documentation shortcomings further complicate debugging, reflecting the current relative immaturity of the quantum software development ecosystem.

## 6.4 Summary

Overall, our findings (summarized in Table 2), suggest that quantum developers primarily rely on traditional, manually intensive debugging strategies, including print statements, code inspection, simulation, visualization, and incremental diagnosis. The use of specialized quantum debugging tools remain rare, with most respondents reporting the use of classical or custom-built solutions. Visualization tools, while helpful in debugging smaller quantum systems, were also noted to scale poorly in more complex scenarios. Although only a minority of developers reported encountering issues with non-determinism, broader challenges—such as long run times, limited hardware access, brittle libraries, and documentation gaps—were frequently cited as factors that impede debugging and prolong bug resolution. These results point to an urgent need for scalable, quantum-specific debugging frameworks that address both the technical and infrastructural limitations faced by quantum developers today.

## 7 RQ3: RECURRING BUGS AND ISSUES FACED BY QUANTUM DEVELOPERS

Quantum developers encounter a range of recurring bugs and issues that differ in both origin and behavior from those typically seen in classical software development. In this context, we define *bugs* as specific defects or errors in the logic, implementation, or expected behavior of quantum programs. We use the term *issues* to refer to a broader category of obstacles, including hardware noise, runtime variability, limitations in development tools or infrastructure [90], as well as developer errors and misunderstandings that may not always manifest as code defects but nonetheless disrupt the development process. Together, these challenges reflect the complexity of the quantum computing stack, the hybrid nature of real-world applications, and the immaturity of current tooling. We present our findings in three parts: first, we describe the most frequently occurring bugs and issues reported by survey respondents; second, we examine their underlying causes as indicated by respondents; and third, we analyze how respondents described these problems as manifesting during the development and execution of quantum applications.

### 7.1 Types of Bugs and Issues Encountered by Quantum Developers

To understand which types of bugs quantum developers encounter most frequently, we asked respondents to estimate the proportion of bugs in their quantum software projects that were quantum-specific, classical, cross-domain, or other (Survey Question B16). Although we recognize that these estimations are subjective and not quantitatively precise, they still offer valuable insight into perceived trends across the hybrid quantum-classical stack. Hybrid quantum-classical applications refer to programs that combine quantum circuits with classical control logic, orchestration code, and data processing routines. Most real-world quantum applications today follow this model, due to the limited capabilities of current quantum hardware and the need to offload tasks to classical systems. The survey responses reflect the practical observations of developers and can help highlight where errors are most commonly encountered in real-world quantum development.

**Bug Frequency by Subsystem.** As shown in Figure 12, classical bugs dominated the responses in terms of frequency (43.5%). These include errors rooted in conventional software development, such as logic flaws, type mismatches, and data handling problems in classical code. Quantum-specific bugs accounted for an average of 34% of bugs based on survey responses. Quantum-specific bugs typically involve issues such as incorrect circuit design, incorrect gate usage, or hardware-induced phenomena such as decoherence or measurement error. Cross-domain bugs, averaging 19% of bugs based on survey responses, refer to faults or problems that span both the classical and quantum layers, sometimes requiring changes in both domains to resolve. Only a small number of developers (two out of 26) indicated that 3.5% or more of their bugs fell into the "other" category. Those respondents cited deployment and configuration issues, including provider API incompatibilities and build-related failures as "other" type bugs. This distribution underscores that, while quantum-specific bugs are prominent, many issues remain rooted in classical software engineering and the broader complexity of hybrid quantum-classical integration. These findings are consistent with previous studies [59, 90]. Specifically, in our previous work analyzing quantum discussion forums, we found that the majority of bugs were classical (43%), followed by quantum-specific (36%), and cross-domain (21%), reflecting comparable trends in bug origin across the hybrid quantum software stack [90].

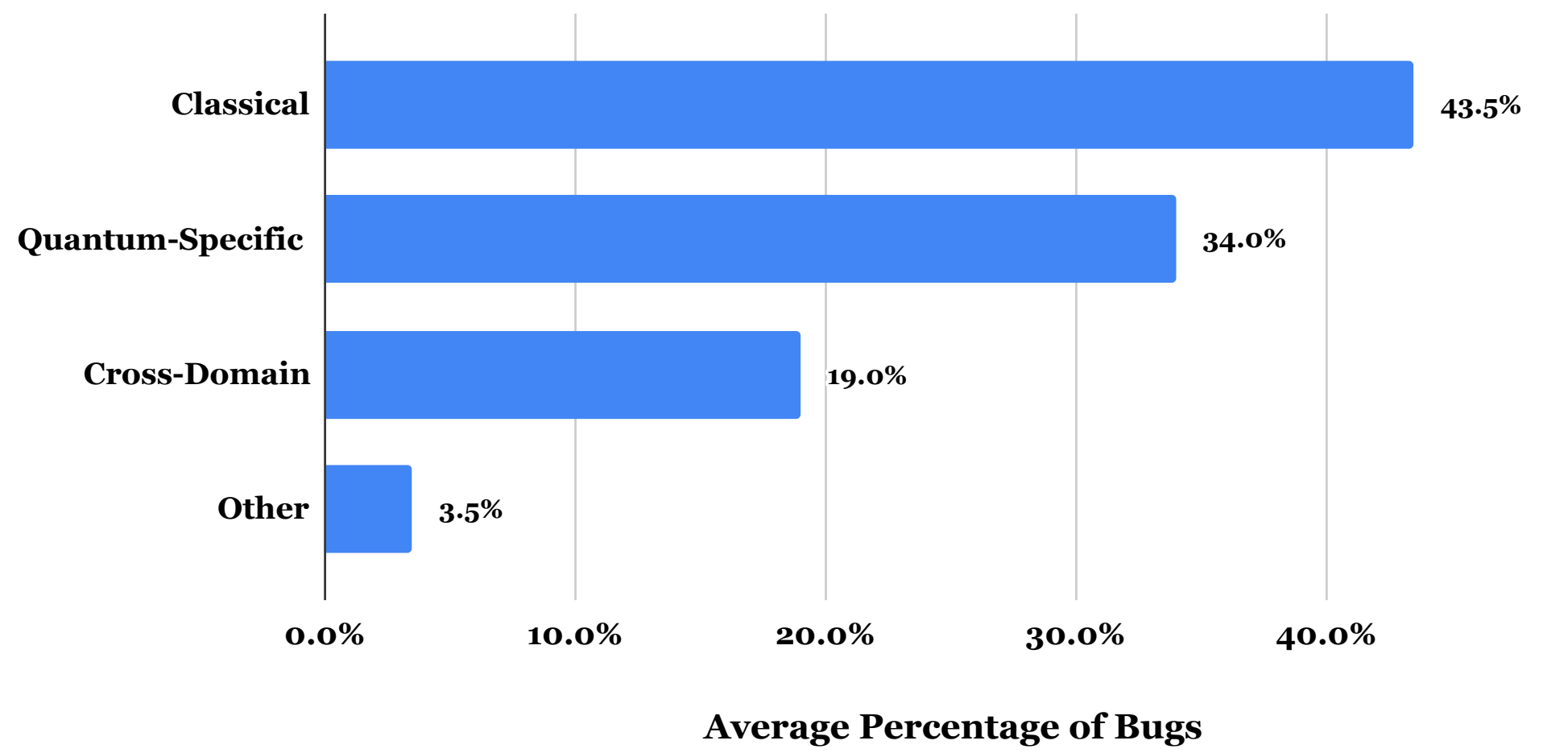

Fig. 12. Percentage of Bugs Encountered by Domain (Survey Question B16)

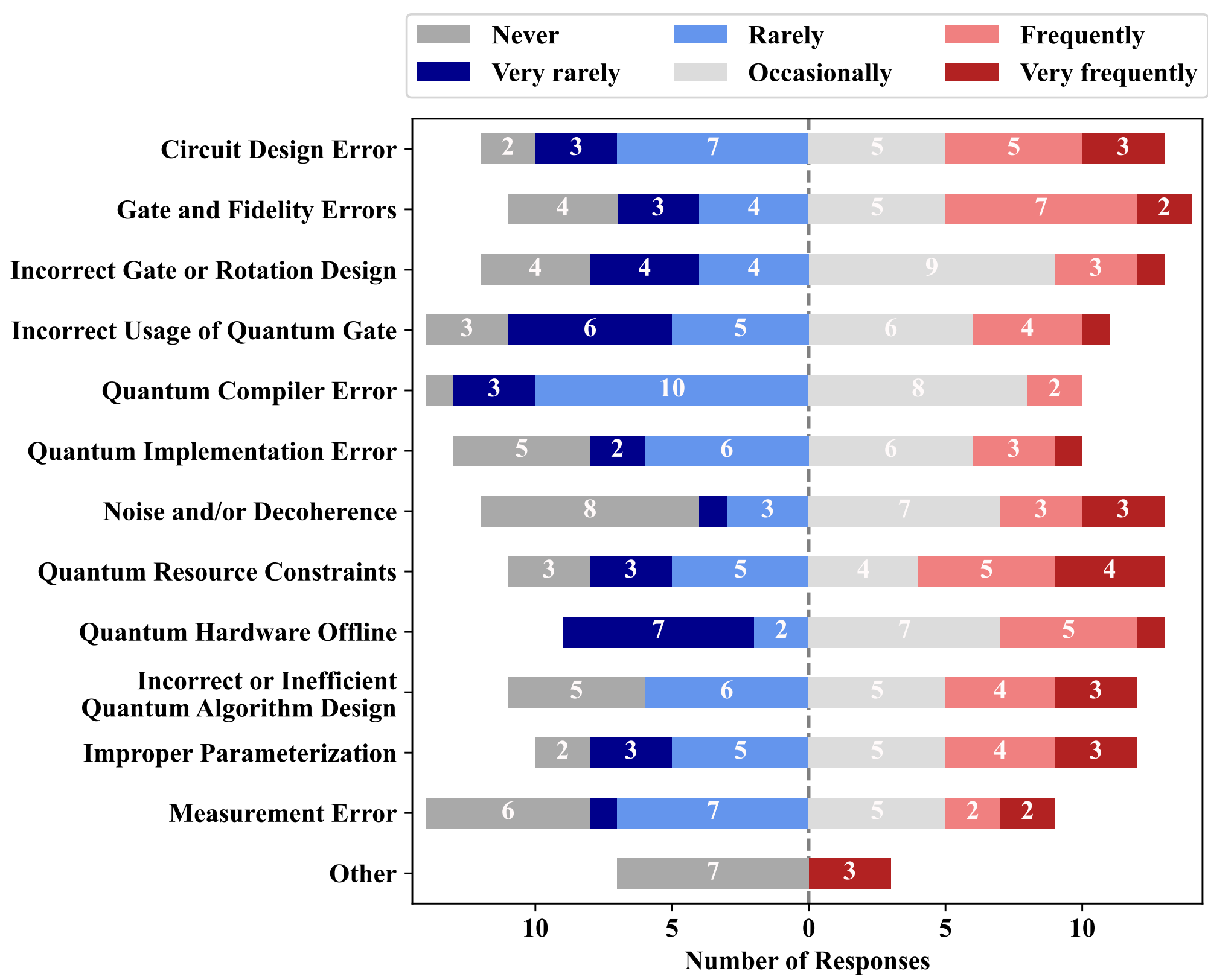

Fig. 13. Frequency of Quantum-Specific Bugs Encountered (Survey Question B17)

**Types of Quantum-Specific Bugs.** In a follow-up question (Survey Question B17), we asked developers how frequently they encounter various categories of quantum-specific bugs. As shown in Figure 13, the most common problems encountered were the following:

- **Gate and Fidelity Errors:** Nine respondents reported frequent or very frequent occurrences.
- **Circuit Design Errors:** Eight respondents also selected frequent or very frequent.
- **Incorrect or Inefficient Algorithm Design:** Seven respondents reported frequent or very frequent occurrences.
- **Improper Parameterization:** Seven respondents selected frequent or very frequent.
- **Noise and/or Decoherence:** Six respondents reported encountering this frequently or very frequently.

Less frequent bugs included hardware offline errors, measurement errors, and compiler-related failures (see Figure 13). These results suggest that developers face the greatest difficulty in areas closely tied to quantum logic and circuit design, where issues are often algorithmic or architectural in nature, as well as in managing the noise and variability inherent in quantum hardware.

It should also be noted that in the open-ended responses to Survey Question B18, two developers raised important concerns about how bugs are framed in the quantum context. One noted that "a lot of the things listed here like 'measurement error' are not bugs, they're just properties of quantum computers," referencing the probabilistic nature of quantum computing and comparing it to how floating-point approximations are treated in classical computing. Another pointed to the deprecated quantum code as a common but frustrating source of error.

These responses highlight that the complexities of QC introduce problems that are difficult to characterize as "bugs" in the classical sense. A measurement error, for instance, could stem from a coding defect in how a measurement is specified, or instead arise from the inherent uncertainty of quantum mechanics or the noise profile of NISQ hardware. The former would align with a traditional notion of a bug, whereas the latter reflects an issue rooted in the probabilistic and error-prone nature of current devices. Quantum developers must therefore remain mindful that many faults encountered in practice may not be attributable solely to code defects, but to the underlying physics and hardware stack.

**Finding 9: Bug Origins Span Classical, Quantum, and Cross-Domain Layers —** Survey respondents reported that bugs in quantum software most frequently originate in classical code (43.5%), followed by quantum-specific code (34%) and cross-domain interactions between classical and quantum components (19%). Among quantum-specific bugs, developers most often encounter issues related to circuit design, parameterization, fidelity, and noise, highlighting the combined impact of algorithmic complexity and hardware variability in NISQ-era systems.

## 7.2 Library and Platform Bugs and Issues

Library- and platform-level issues are a well-documented source of friction in quantum software development. Previous studies have shown that these bugs occur frequently, often due to broken dependencies, the rapid evolution of quantum libraries, and poorly documented changes, which are recurring factors that disrupt quantum workflows [14, 50, 90]. Due to the prevalence and impact of these issues, we explicitly included questions about library and platform bugs in our survey to better understand their effect on real-world development.

We asked how often bugs or errors occur after a library or platform update (Survey Question L3). As shown in Figure 14, 21 of the 26 respondents reported encountering such issues at least occasionally, with 10 selecting *occasionally*, eight selecting *frequently*, and three selecting *very frequently*. Only four respondents reported such bugs as *rare*, and just one indicated they had *never* experienced issues after an update. These findings indicate that platform and library updates are a pervasive source of instability in quantum software development—often attributed to dependency conflicts, backward compatibility breaks, and API deprecations—and echo prior studies documenting similar disruptions, particularly in fast-evolving frameworks like Qiskit [83, 90]. Although continuous updates are essential to progress, the high rate of regressions suggests that backward compatibility, dependency management, and release testing remain underdeveloped across much of the quantum ecosystem. Consequently, many developers may treat updates as risky operations to be delayed or approached with caution, increasing their maintenance burdens and hindering their development workflow.

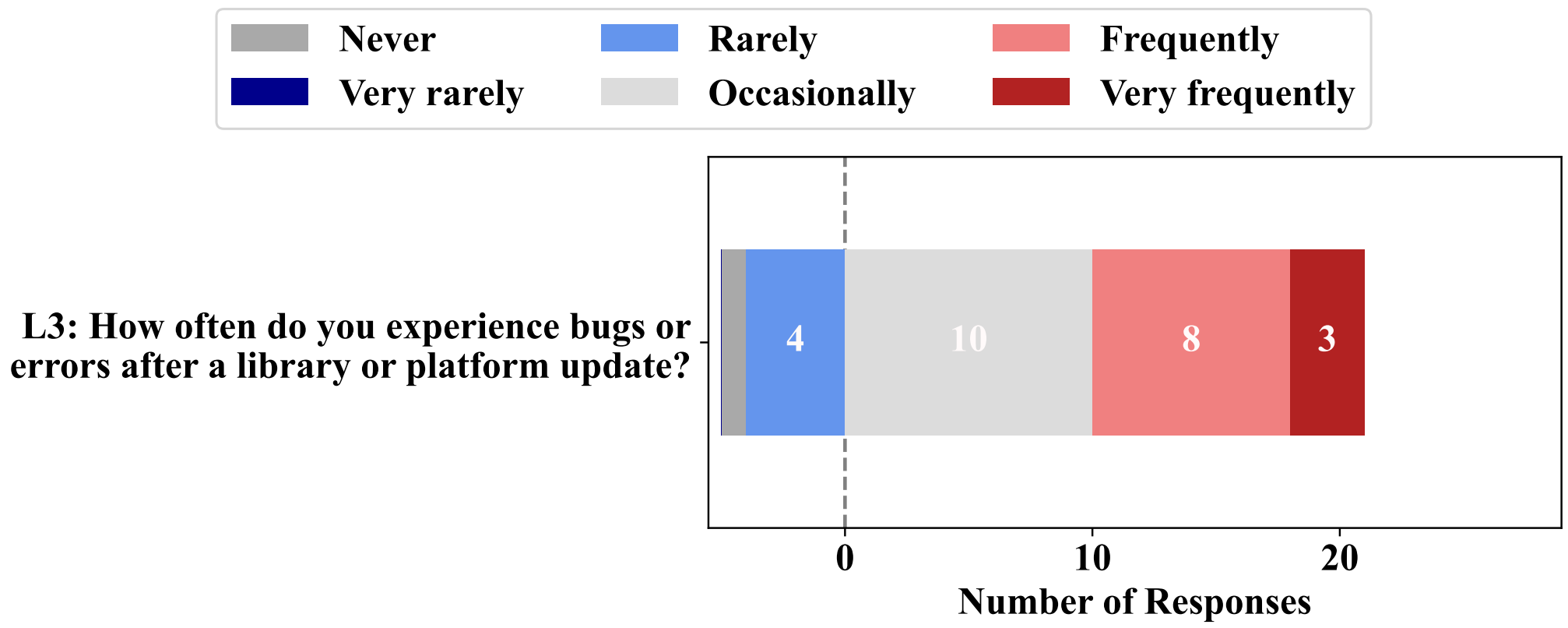

Fig. 14. Platform and Library Issues Encountered by Quantum Developers (Survey Question L3)

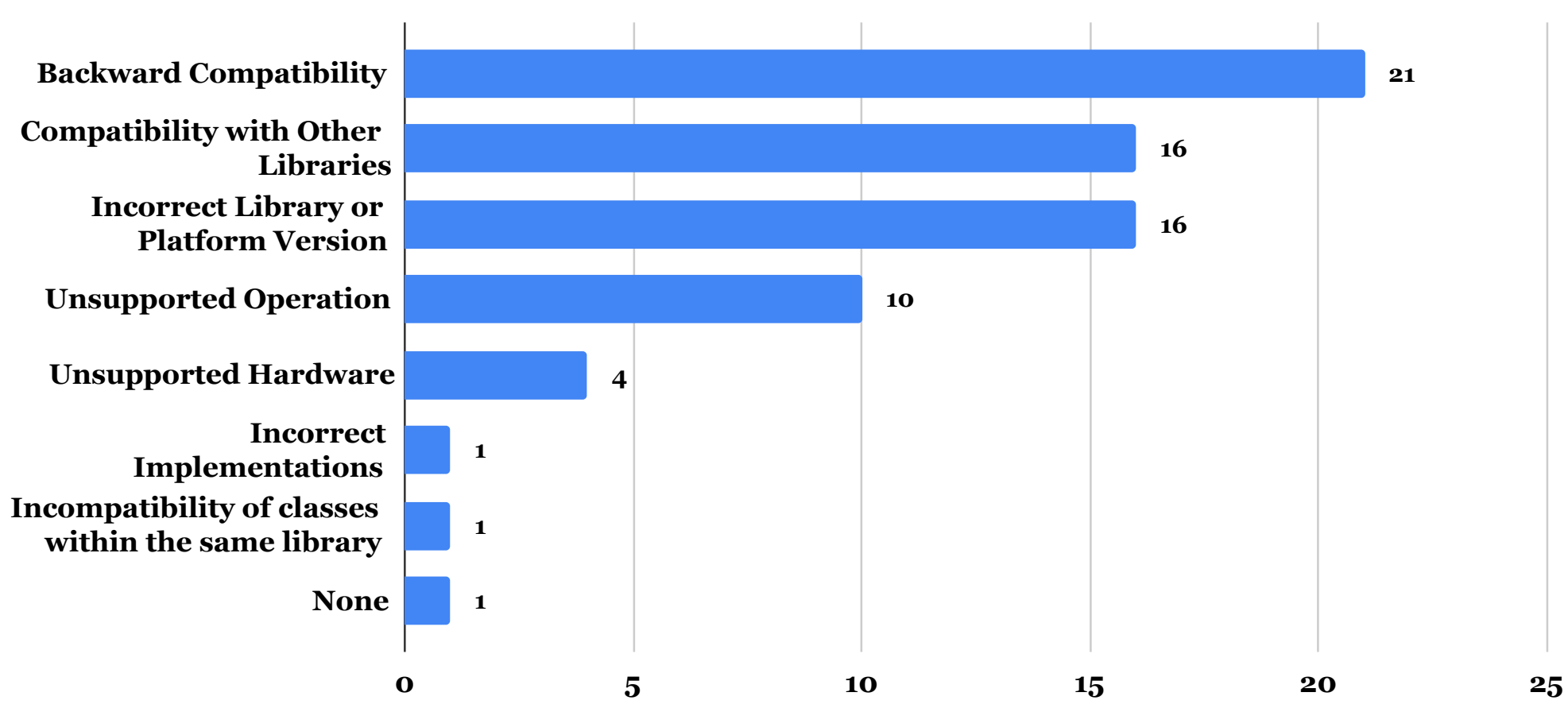

Fig. 15. Platform and Library Issues Encountered by Quantum Developers (Survey Question B15)

To capture the range of library- and platform-related issues encountered by quantum developers, we provided a list of common bug categories and asked respondents to select all that they had experienced (Survey Question B15). As shown in Figure 15, the most frequently selected issue was *backward compatibility* (21 respondents), followed closely by *compatibility with other libraries* and *incorrect library or platform versions*, each selected by 16 respondents. *Unsupported operations* were noted by 10 respondents, and *unsupported hardware* by four. The less frequently selected categories included *incorrect implementations* and *incompatibility within the same library* (one selection each). Only one participant reported that they did not encounter any of the listed platform and library issues provided. These results indicate that library and platform bugs are not only widespread and commonly encountered by quantum developers, but also span a diverse set of technical challenges—including versioning, interoperability, hardware compatibility, and internal consistency.

Additionally, to better understand how developers manage changes in their quantum development environments, we asked a series of questions aimed at assessing update practices, awareness of library and platform changes, and whether such updates are more likely to introduce or resolve bugs. First, we ask about their update practices (Survey Question L1) as it relates to quantum libraries and platforms. The respondents were split. Specifically, six participants said they update libraries as soon as new versions are available, another six reported doing so only when updating their runtime environments, and six said they update sporadically. Only three followed a regular schedule. Notably, five respondents indicated they update libraries specifically in response to encountering bugs, suggesting that update-driven instability is a known risk that some developers try to avoid until absolutely necessary.

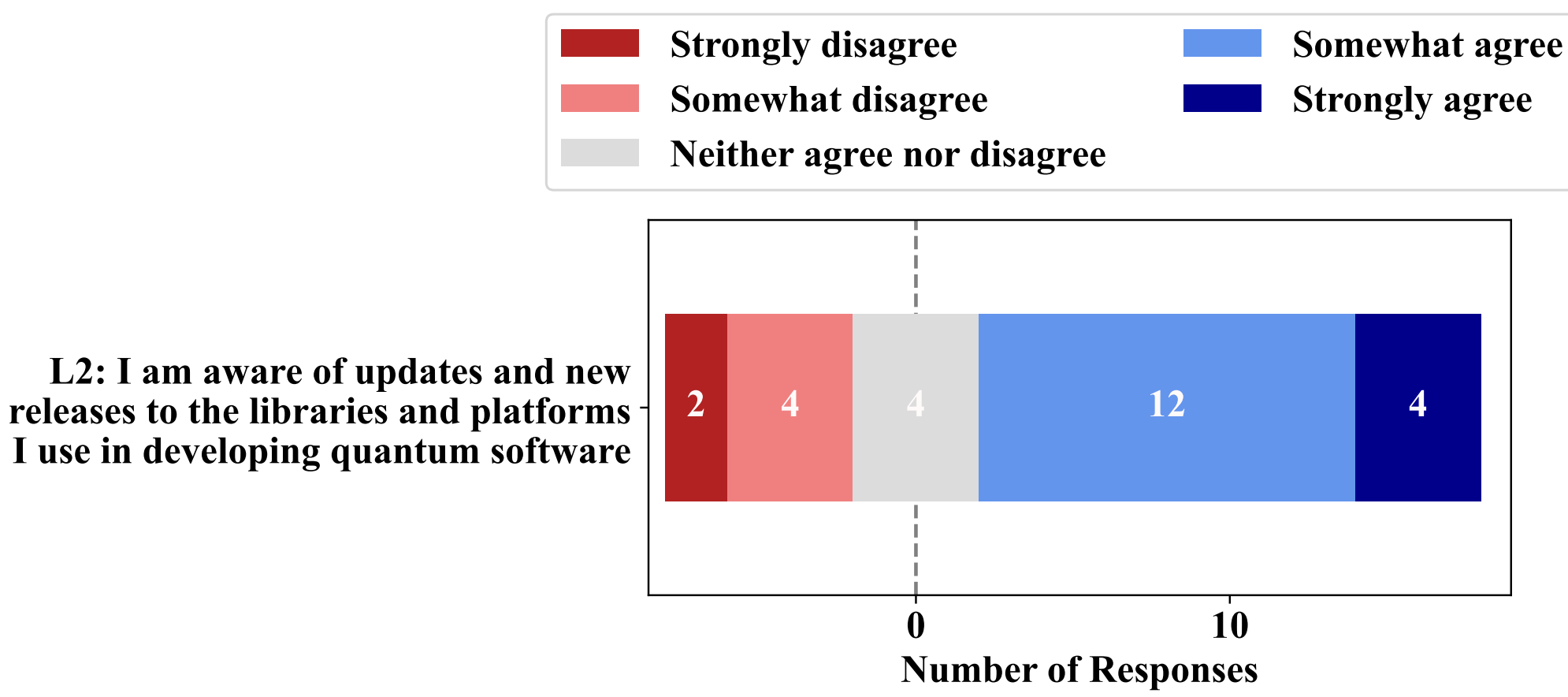

Fig. 16. Awareness of Quantum Platform and Library Releases and Updates (Survey Question L2)

Survey Question L2 asked respondents whether they felt aware of new releases or updates to the quantum libraries and platforms they use. As shown in Figure 16, while 16 respondents agreed or somewhat agreed, six out of 26 were neutral or disagreed. These results suggest that, despite overall moderate awareness, a significant subset of developers may struggle to keep up with updates—potentially due to inconsistent release communication, lack of centralized changelogs, or limited community outreach by library maintainers.

These themes were reinforced in open text responses to Survey Question L4, which focused on on bugs encountered by developers in libraries and platforms. A respondent observed that "Qiskit updates have caused the most problems," adding that if they were evaluating update-related bugs solely based on their experience with Qiskit, they would have selected "very frequently" in Survey Question L3. Another noted that "when Qiskit was upgraded, everything changed" and expressed frustration with the lack of a smooth migration path. A third respondent described the transition to Qiskit 1.0 as "quite impractical," emphasizing that "almost the entire API changed." Notably, Qiskit deprecated several libraries on its platform in 2021 and 2022, no doubt causing headaches for many developers [69]. Finally, a fourth respondent cited general problems with "dependency issues and conflicts and deprecations" in quantum platforms and libraries. These responses illustrate the disruptive potential of rapid platform evolution, particularly when combined with insufficient backward compatibility, documentation gaps, or ecosystem fragmentation.

Together, these findings point to a fragile and often frustrating update experience for many quantum developers. Frequent regressions, breaking changes, and poor migration support—particularly in widely used rapidly changing platforms like Qiskit—highlight the need for more stable release cycles, clearer communication with developers, and stronger tooling to support version management in the quantum software ecosystem.

**Finding 10: Library and Platform Updates Introduce Frequent Breakages —** Over 80% of respondents reported encountering bugs following library or platform updates at least occasionally, with 42% experiencing such issues frequently or very frequently. These problems were most often attributed to backward compatibility breaks, dependency conflicts, and version mismatches. Open-ended responses further highlighted significant developer frustration with major frameworks such as Qiskit, where abrupt API changes and limited migration support frequently disrupt development workflows.

## 7.3 Causes of Recurring Bugs and Issues

Understanding the underlying causes of bugs is essential for building more reliable quantum software systems. Survey Questions B11-B13 explored the sources of recurring bugs as well as the contributing factors behind developer errors. The findings echo themes from our prior work [90], while also highlighting ongoing pain points in the quantum ecosystem.

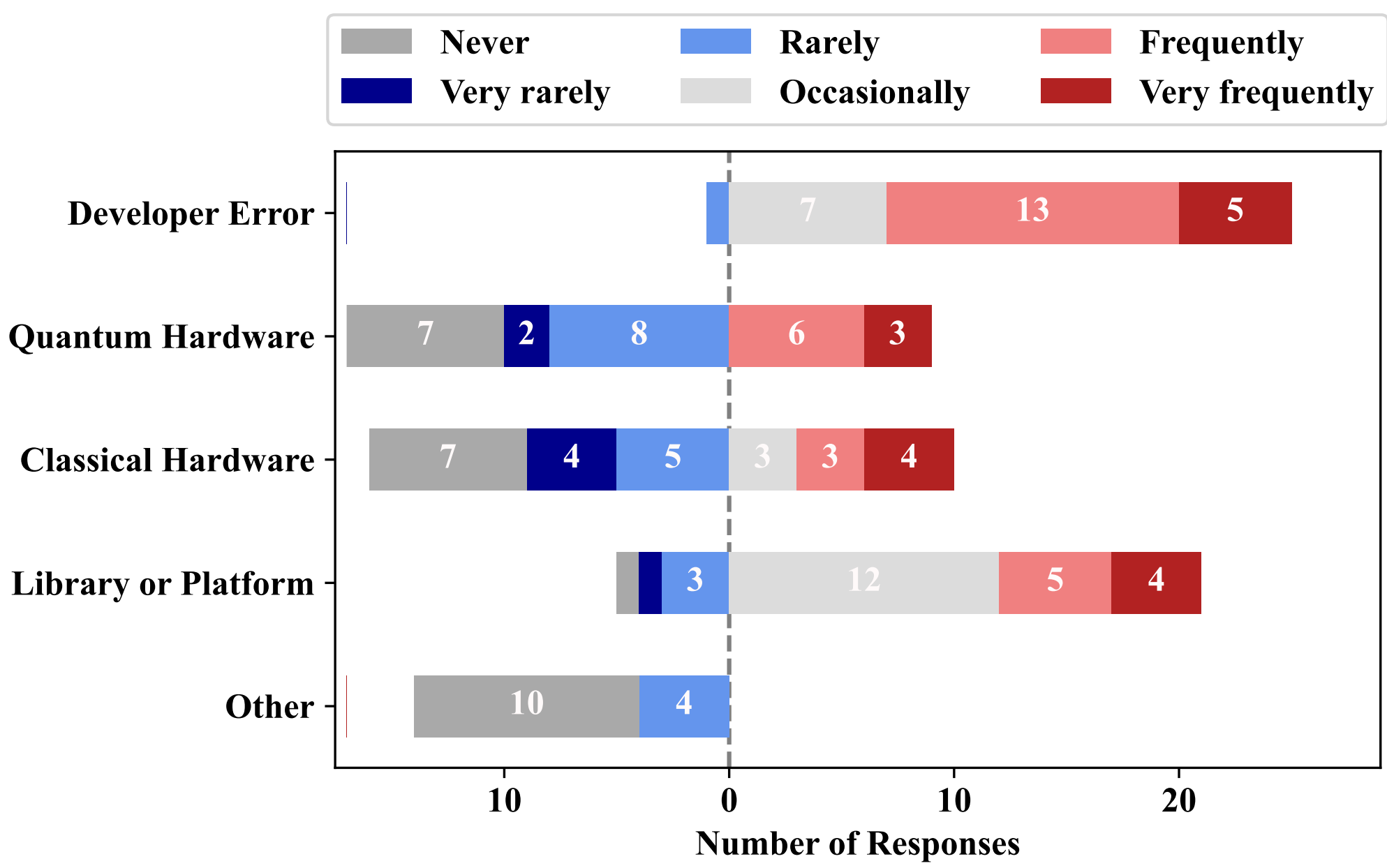

Fig. 17. Causes of Quantum Bugs and Issues (Survey Question B11)

*7.3.1 Developer Error.* Developer errors emerged as the most frequently cited cause of bugs. As shown in Figure 17, 18 out of 25 respondents reported that developer errors were the cause of bugs *frequently* or *very frequently*. Here, we use the term "developer errors" broadly to encompass both low-level mistakes—such as conceptual misunderstandings or oversights—and concrete code defects, such as mis-specified gates or incorrect parameter values. These errors introduce faults into the code that may later manifest as run-time errors or failures during execution. Examples of developer errors include logic flaws in classical control code, incorrect gate or parameter usage in quantum circuits, and broader misunderstandings of quantum behavior that can misguide circuit or algorithm design. As Interview Respondent 1 explained, "You need to do a lot of thinking in how to translate bit logic to the so-called qubit logic," emphasizing the conceptual gap developers must bridge when writing quantum software. Prior studies have similarly found that the abstract and non-intuitive nature of quantum computation makes it particularly susceptible to developer errors [90].

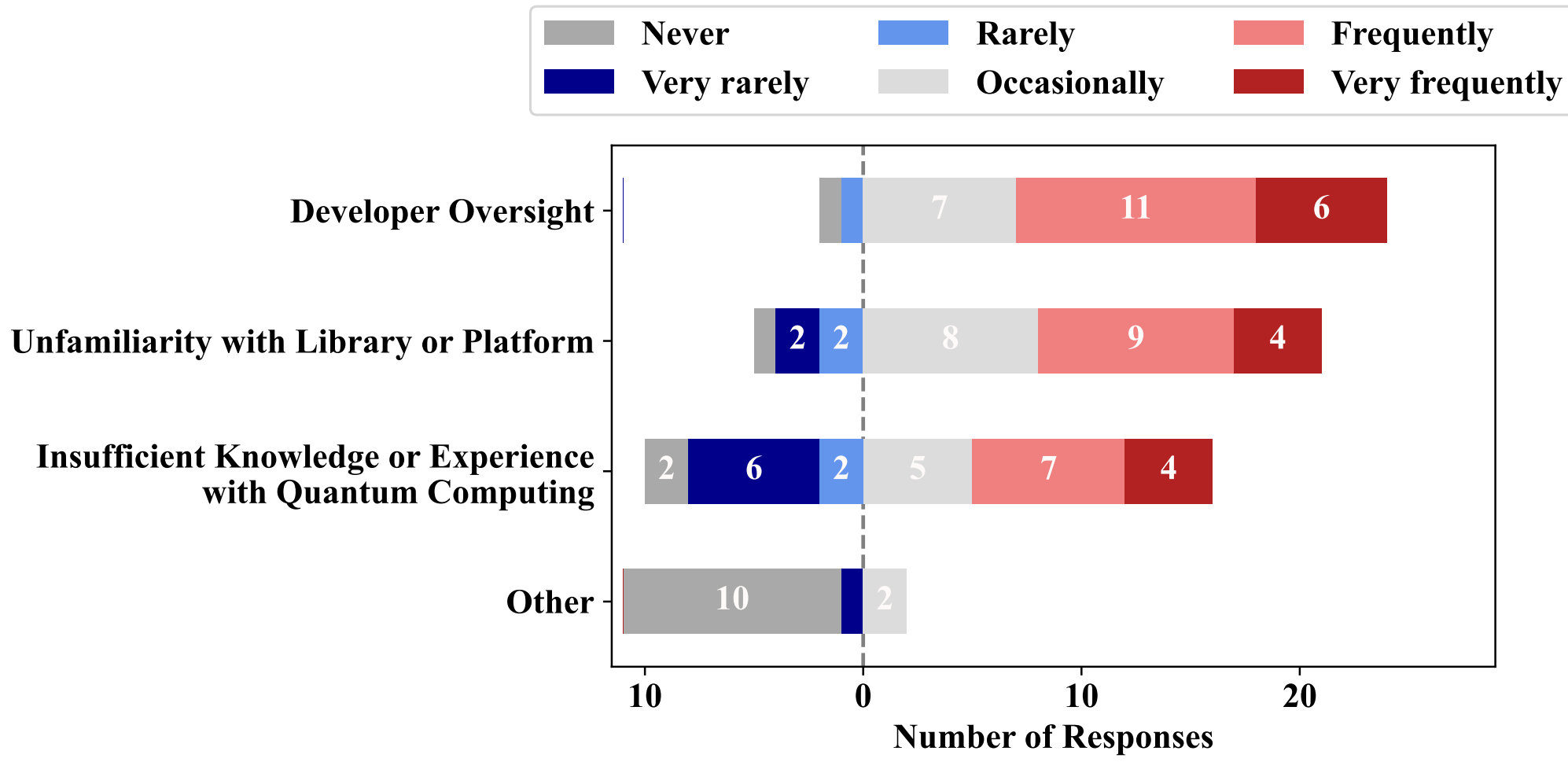

Fig. 18. How Bugs Are Introduced in Quantum Applications (Survey Question B13)

Responses to Survey Question B13 shown in Figure 18 further confirmed this trend. Specifically, 17 respondents reported that developer oversight occurs *frequently* or *very frequently*, while 13 cited unfamiliarity with the platform or libraries as a cause of faults. Additionally, 11 respondents indicated that insufficient quantum knowledge also plays a substantial role in triggering bugs. Taken together, these results indicate that developer errors in quantum software are driven not merely by oversight, but by deeper challenges related to limited familiarity with quantum platforms, insufficient quantum domain knowledge, and the difficulty of reasoning at low levels of abstraction without adequate tooling support.

**Finding 11: Developer Errors Are the Leading Cause of Bugs —** Sixty-nine percent of respondents reported that developer errors contribute to bugs frequently or very frequently, citing oversights, unfamiliarity with libraries or platforms, and gaps in quantum-specific knowledge as substantial contributing factors.

*7.3.2 Quantum and Classical Hardware.* In contrast to software-related sources, hardware—both quantum and classical—was less frequently cited as a root cause of bugs. Only nine respondents reported that quantum hardware was a *frequent* or *very frequent* source of bugs, while only seven said the same about classical hardware (see Figure 17). Reported quantum hardware issues included difficulties accessing quantum devices through provider APIs, long queue times, and inconsistent execution results across different machines. For classical hardware, several respondents pointed to slow execution times on simulators—especially when simulating larger circuits—as well as incompatibilities with specific hardware configurations, such as GPUs. Taken together, these findings suggest that while hardware constraints introduce friction into development workflows, the majority of bugs originate in higher layers of the stack, including application logic, cross-domain integration, and platform-level instability.

Notably, two developers pointed out that bugs often emerge at the interface between software and hardware. These include issues such as misconfigured runtime environments or failures in resource-intensive orchestration logic. Such problems blur the boundary between software and hardware, making them especially difficult to diagnose and resolve.

**Finding 12: Hardware-Related Bugs Are Reported Less Frequently —** Hardware was cited less often than software as a frequent source of bugs, with 35% of respondents reporting quantum hardware and 27% reporting classical hardware as frequent or very frequent contributors.

*7.3.3 Other Causes and Observations.* Two respondents cited other causes of bugs outside the primary categories. One noted that errors often stem from incorrectly specified input data, particularly when complex states in superposition are involved: "While the quantum software works as expected, it still can be difficult to find an error in the used input data (for example, a complex input state in superposition)." The other cited inexperience with the programming language being used as a source of error, emphasizing the importance of developer familiarity with tools and syntax.

Taken together, these findings illustrate the multifaceted nature of bug origins in quantum software. While developer errors remain the most frequently cited source, often exacerbated by the steep learning curve of quantum computing, bugs can also arise from the fragility of platform dependencies, gaps in abstraction, and subtle interface mismatches between hardware and software. These insights reinforce the need for improved developer tooling, clearer abstractions, and better communication of platform changes to reduce the frequency and impact of errors in future quantum systems.

## 7.4 Bug and Issue Manifestations

While the previous section explored the causes of quantum software bugs, it is equally important to understand how bugs and other issues manifest in quantum software. Survey Questions B1–B10 examined this topic by asking developers how frequently they observe manifestations such as crashes, incorrect outputs, warnings, and other signals of failure. The findings support themes from our prior research [90], while providing updated insights into how errors present themselves in practice.

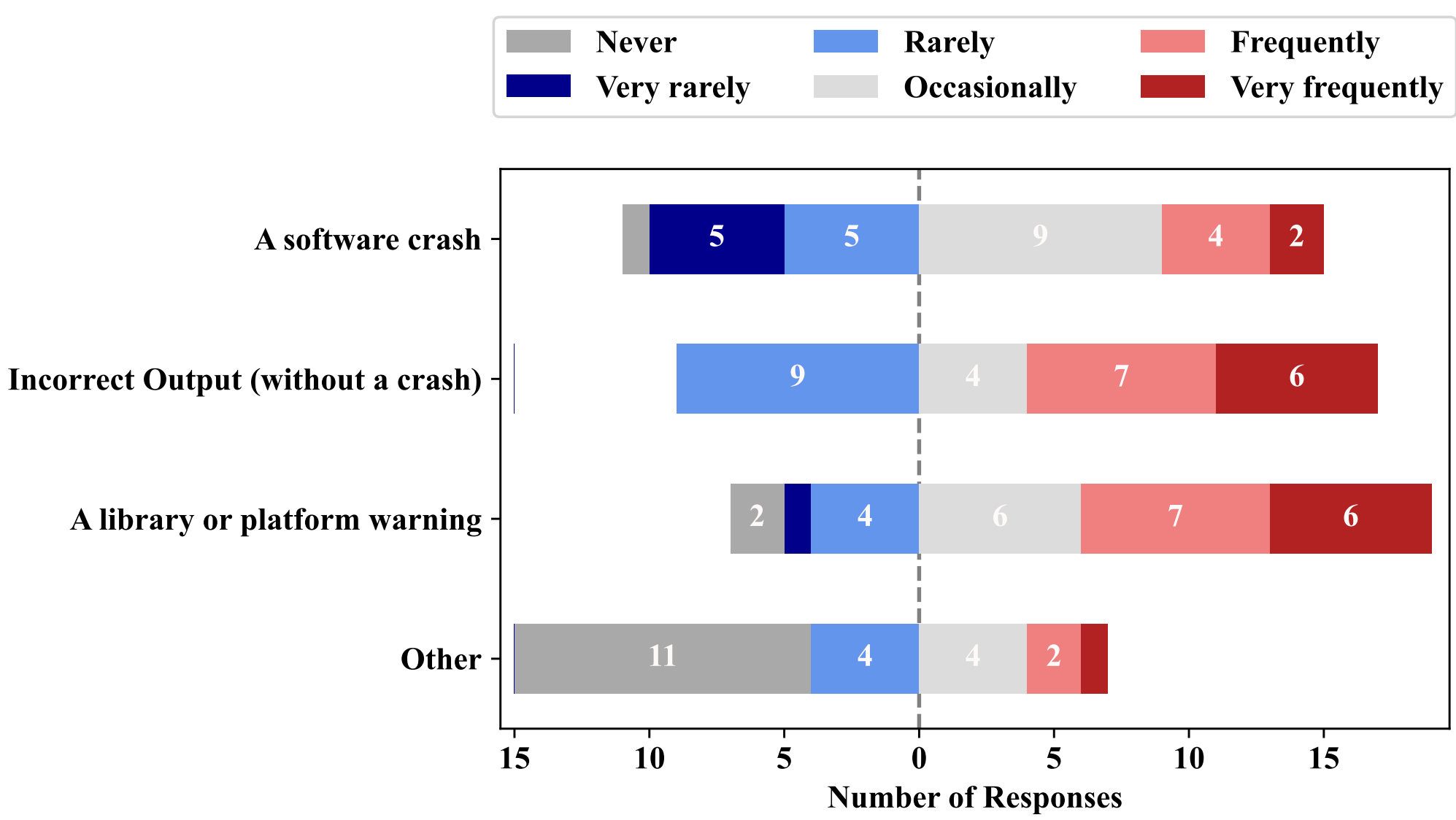

Fig. 19. Bug Manifestations in Quantum Software (Survey Question B1)

*7.4.1 Incorrect Output.* The most common manifestation of bugs was *incorrect output without crashing*. As shown in Figure 19, 13 respondents reported encountering incorrect application output *frequently* or *very frequently*, making it—along with warnings, discussed below—the most widely observed manifestation across all categories. These "silent" failures are particularly problematic in quantum computing, where probabilistic outputs may appear plausible even when incorrect. As Interview Respondent 3 noted, “You can’t always trust the output. Sometimes it looks fine, but it’s not what you expect when you analyze it deeply.”

Additional responses to Survey Question B6 further illustrate this challenge. One participant explained that “incorrect preparation of state [and] influence of noise” were common sources of unexpected behavior or incorrect application output. Another respondent reported that “inputs where the shape or order of the parameters produced a very misleading output” often led to subtle bugs. These examples highlight how incorrect outputs can arise from low-level configuration issues or physical effects, making them difficult to detect without repeated trials, statistical checks, or deep domain expertise.

**Finding 13: Incorrect Output Without Crashes Is the Most Common Bug Manifestation —** Thirteen out of 26 respondents (50%) reported frequently encountering bugs that produce incorrect outputs without crashing, creating silent failures that are particularly difficult to detect and diagnose in probabilistic quantum applications.

*7.4.2 Crashes.* *Crashes* were less frequently reported than incorrect outputs or warnings, but they still represented a meaningful class of bug manifestations. According to Survey Question B1, six of 26 respondents indicated that software crashes occurred either *frequently* or *very frequently* in their quantum software projects. These crashes typically arose during simulation or in the classical orchestration layer, rather than from execution on the quantum backend. As Interview Respondent 4 remarked, “The classical side breaks before the quantum side ever does,” highlighting a recurring point of failure in hybrid applications.

Insights from Survey Question B3 further illuminate the root causes: one participant explained, “It’s mostly due to memory limitations,” while another pointed to infrastructure issues such as “build and configure complications, with config or build error messages, or link-time issues, that present as completely unrelated to the culprit library.” A third respondent highlighted cross-library incompatibilities, noting that they encountered issues “whenever there is an incompatibility between Qiskit and other Python libraries.” That respondent went on to give an example: “Google Collaborate doesn’t install Qiskit by default, which is very annoying.” Together, these responses suggest that crashes may more often stem from low-level technical constraints—such as resource limits, tooling mismatches, and environmental inconsistencies—rather than from faults in quantum logic or errors in quantum algorithm implementation.

**Finding 14: Crashes Are Often Linked to Resource and Environment Issues —** Six out of 26 respondents (23%) reported frequent crashes, most often stemming from memory limitations, build or configuration errors, and cross-library incompatibilities—typically occurring during classical orchestration or simulation of quantum applications, rather than during execution on quantum hardware.

*7.4.3 Warnings.* *Warnings* emerged as a leading manifestation of bugs, with 13 respondents reporting *frequent* or *very frequent* occurrences. According to participants, these typically stemmed from SDK changes, deprecated features, or shifts in backend behavior. Developers noted that warnings, while not fatal, could alter program behavior in subtle ways. One respondent noted that “deprecation warnings [and] memory warnings” often surfaced during development, while another described how “we see a constant stream of NumPy warnings from tools like Qiskit,” which they felt diminished the overall quality of the platform. Another participant emphasized the disruptive role of breaking API changes and update churn: “Libraries update a lot, often breaking interfaces (or warning of soon to be broken ones).” Additional concerns included “lots of C/C++ warnings in things like XACC and CUDA Quantum, many of which are signals of numerical problems,” and “Qiskit 1.0 is not fully back compatible with 0.45 version.” Collectively, these responses underscore how even non-fatal warnings can introduce ambiguity and risk, particularly in ecosystems that evolve rapidly and lack consistent deprecation or migration guidance.

**Finding 15: Warnings Are a Common but Ambiguous Bug Manifestation —** Half of respondents (13 out of 26) reported frequently encountering warnings during quantum software development. While non-fatal, these warnings often stem from SDK changes, deprecated features, or backend updates, and were perceived as introducing ambiguity, behavioral risk, and reduced confidence in platform stability.

*7.4.4 Other Manifestations.* A smaller number of respondents (seven out of 25) identified other types of bug manifestations not captured by crashes, output errors, or warnings. Respondents reported these manifestations included jobs becoming stuck in execution without progress, jittery or non-converging outputs in variational quantum algorithms, and the complete absence of output or diagnostics. While some of these behaviors—such as infinite loops—are also common in classical software, others are more reflective of quantum-specific phenomena. For example, variational algorithm instability may result from a complex interplay of noise, improper parameter tuning, or optimization landscape characteristics that are unique to NISQ-era devices. One participant emphasized the difficulty of identifying such issues in a maturing ecosystem: "Most of the bugs come from poor testing of Python code... the community is still developing." These diverse manifestations reinforce the broader challenge of fault diagnosis in quantum software, where bugs may not surface through conventional failure modes, and the lack of consistent runtime feedback or tooling makes isolation and resolution particularly difficult.

**Finding 16: Some Bug Manifestations Fall Outside Classical Categories —** Seven out of 26 respondents (27%) reported bugs manifesting in non-standard ways, including job hangs, output jitter, and missing results, underscoring the challenges of diagnosing faults in an evolving and unpredictable quantum software environment.

### 7.5 Summary of Recurring Bugs and Issues

Quantum software development presents a wide spectrum of recurring bugs rooted in both classical and quantum domains. While classical bugs remain the most frequently encountered, developers also reported quantum-specific issues tied to circuit design, noise, and parameter tuning, as well as cross-domain bugs that span both subsystems.

Library and platform bugs are a major source of instability, often triggered by breaking changes, dependency conflicts, and insufficient migration support—especially in widely used frameworks like Qiskit. Developer errors, driven by oversight, platform unfamiliarity, and the complexity of quantum logic, emerged as the leading root cause of bugs.

Finally, bugs in quantum systems frequently manifest in non-obvious ways—such as incorrect output, subtle warnings, or stalled jobs—making diagnosis and resolution particularly difficult. These findings underscore the need for better abstractions, more stable tooling, and improved support for identifying and managing quantum-specific errors. Table 2 presents the key findings with respect to RQ3.

## 8 DISCUSSION AND IMPLICATIONS

The results of our study offer a detailed look at the real-world practices and challenges faced by quantum software developers. Although some findings reinforce the trends observed in earlier studies, our data also highlight emerging concerns shaped by the growing complexity and scale of quantum applications. In this section, we reflect on the implications of our findings, discuss their relevance for future tool development and research, and identify key areas where QSE must mature to better support developers.

| **Key Findings for RQ1: Testing Practices and Challenges** |
|---|
| **F1** Most respondents relied on traditional testing methods, with 23 out of 26 (88%) reporting unit testing, 17 (65%) reporting integration testing, and 14 (54%) each reporting regression and acceptance testing. |
| **F2** Only eight out of 26 respondents (31%) reported using quantum-specific testing tools, reflecting limited adoption of such tools. |
| **F3** Approximately two-thirds of respondents relied on *ad hoc* and manual testing practices, including handcrafted assertions and visualization, while also reporting challenges such as flaky tests and scalability limitations in visualization. |
| **F4** Eleven out of 26 respondents (42%) identified non-determinism as a key testing challenge, alongside noise and machine-to-machine variability. |
| **Key Findings for RQ2: Debugging Practices and Challenges** |
| **F5** Eighteen respondents (69%) reported relying on diverse, but largely manual debugging strategies, often without the use of specialized quantum debugging tools. |
| **F6** Very few participants reported using tools designed specifically for quantum debugging, with most instead relying on classical or general-purpose tools. |
| **F7** Fourteen respondents (54%) indicated that they used circuit or state visualizations during debugging, though many noted that these approaches did not scale effectively for larger systems. |
| **F8** Respondents highlighted that bug resolution was often hampered by long execution times, limited hardware access, and frequent library or dependency issues, compounded by poor library documentation. |
| **Key Findings for RQ3: Bug and Issue Types, Causes, and Manifestations** |
| **F9** Respondents estimated that bugs originated on average 43.5% from classical code, 34% from quantum-specific code, and 19% from cross-domain interactions, indicating that failures frequently arise across multiple layers of quantum systems. |
| **F10** More than 80% of respondents reported encountering bugs due to library or platform updates at least occasionally, with 42% experiencing them frequently or very frequently; common issues included backward compatibility breaks and dependency conflicts. |
| **F11** Developer errors were the leading cause of bugs, with 18 out of 25 respondents reporting them as frequent or very frequent; oversight, unfamiliarity with platforms or libraries, and insufficient quantum knowledge were substantial contributing factors. |
| **F12** Hardware-related bugs were less commonly reported, with 35% of respondents citing quantum hardware and 27% citing classical hardware as frequent sources of bugs. |
| **F13** Incorrect output without program crashes was the most common manifestation of bugs, reported by 13 out of 26 respondents (50%). |
| **F14** Crashes were reported as a manifestation of bugs by six respondents (23%), most often stemming from memory limitations, build or configuration errors, and cross-library incompatibilities in classical orchestration or simulation stages. |
| **F15** Warnings were a common but ambiguous manifestation of bugs, with 13 respondents (50%) reporting frequent occurrences tied to SDK changes, deprecated features, or backend updates. |
| **F16** Seven respondents (27%) reported bugs manifesting in non-standard ways, including job hangs, output jitter, or missing results, reflecting the evolving and unpredictable nature of quantum software debugging. |

Table 2. Key findings of our study for each research question (RQ1–RQ3).

### 8.1 Challenges Unique to Quantum Software Engineering

Our results reinforce that quantum development is not merely classical development augmented with a quantum library—it is a distinct discipline shaped by unique constraints and cognitive demands. While some classical practices can be adapted, QSE introduces novel challenges rooted in the fundamental principles of quantum mechanics, the complexities of hybrid architectures, and the immaturity of current tools and ecosystems.

First, quantum programs are inherently probabilistic. Unlike deterministic classical code, quantum circuits generate distributions of outcomes—even when functioning correctly. As noted in Section 7.4, more than half of respondents (65%) reported encountering incorrect output without accompanying errors, a manifestation that often requires statistical reasoning or repeated executions to identify. This makes even basic correctness checking significantly more challenging.

Second, quantum programs suffer from limited observability. Due to measurement collapse, developers cannot directly inspect program state during execution. Traditional debugging methods such as setting breakpoints or performing step-by-step inspections are not feasible. As a result, many developers (54%) rely on visualization tools or simulators (see Section 6.1), despite their limitations in scalability and their inability to fully emulate the noise characteristics of real hardware.

Third, hybrid quantum-classical architectures introduce additional layers of complexity in testing and debugging. Several respondents reported crashes and failures occurring during the coordination of quantum and classical components—especially in pre- and post-processing classical logic—underscoring the fragility of the toolchain and the difficulty of diagnosing cross-layer issues (see Section 7.1). These challenges are further exacerbated by discrepancies between simulated outputs and the behavior of actual quantum hardware, which tend to grow more pronounced at scale.

Finally, abstraction boundaries in quantum development remain poorly defined. Developers are often required to manage low-level circuit implementation, classical orchestration, and hardware-specific constraints simultaneously. This cognitive burden contributes to the high rate of developer-introduced bugs (65%) observed in our results (Section 7.3).

Together, these findings affirm that QSE requires rethinking established models of correctness, observability, and abstraction. Addressing these challenges will demand the development of new tools and practices tailored to the distinct realities of quantum computing—rather than extensions of classical paradigms.

### 8.2 Limited Adoption of Testing and Debugging Tools

Although quantum developers routinely engage in testing and debugging, our findings reveal that these practices are constrained by several structural and technical barriers. Developers are often forced to rely on manual methods and classical software engineering strategies. This is not because those approaches are ideal, but because the quantum tooling ecosystem lacks mature, integrated support for systematic testing and debugging.

One significant barrier is the limited adoption and visibility of quantum-specific testing tools. Although various academic tools have been proposed—such as Muskit [52], QuCAT [85], and QmutPy [24]—only a very small number of our survey respondents reported using them (see Section 5.2 and Section 6.1). Based on survey responses, only a few were aware that such tools exist. This disconnect between tool development and industry adoption may reflect barriers such as limited outreach, inadequate documentation, or lack of integration with production-ready frameworks.

Even utilities bundled with widely used quantum platforms see limited use as reported by survey respondents. For example, only four out of 26 participants reported using Qiskit Test Utilities, despite their inclusion in the free open-source Qiskit SDK. Several developers cited insufficient flexibility, poor scalability, and a lack of usability as barriers to adoption. Others noted that they resorted to developing in-house testing mechanisms better suited to their specific workflows and environments.

These findings underscore that, while there is clear interest in testing and debugging among practitioners, existing tools often do not meet developers' practical needs. Adoption remains low—only 31% of respondents reported using quantum-specific testing tools—and even when such utilities are bundled with widely used SDKs [70], participants described barriers to their effective use. Academic tools such as MorphQ, Muskit, QuCAT, and QMutPy [24, 52, 60, 85] face even steeper hurdles: Many practitioners were unaware of them and those who were often viewed them as research prototypes lacking integration with production workflows. Bridging this gap will require closer collaboration between tool developers and practitioners, as well as a focus on integration-first design, scalability beyond toy examples, improved usability and documentation, and participatory approaches that align features with developer needs. Greater alignment between research efforts and day-to-day workflows will be essential for widespread adoption.

### 8.3 Barriers to Effective Testing and Debugging

*8.3.1 Low-Level Abstractions.* A major barrier to effective testing and debugging of quantum applications is the low level of abstraction offered by quantum SDKs. Developers are frequently required to manually manage gate sequences, qubit assignments, and backend-specific constraints—challenges echoed in our findings from Section 5 and Section 6, where visual inspection manual trial-and-error, and simulator-based validation emerged as common testing and debugging strategies. These low-level methods increase the likelihood of implementation errors and complicate debugging. As Interview Respondent 1 explained, "You need to do a lot of thinking about how to translate bit logic to the so-called qubit logic," emphasizing that the translation between classical intent and quantum execution is nontrivial. The absence of robust language features, static analysis, or type checking (a result of most quantum libraries being implemented in Python) further compounds this challenge, leaving developers without guardrails for catching simple mistakes early.

*Mitigation strategies.* One way to reduce the cognitive burden of low-level circuit construction is through higher-level abstractions and language features that catch mistakes earlier. Respondents noted the challenges of reasoning directly at the gate and qubit level, which increases the likelihood of errors such as mis-specified gates or incorrect qubit indices. While few such tools exist today, more structured programming models, stronger type systems, circuit templates, the integration of compile-time checks, and more mature static analysis tools would help mitigate these issues by catching defects earlier and reducing reliance on manual inspection and trial-and-error reasoning.

*8.3.2 Library and Platform Instability.* Library and platform instability also contributes to the complexity of testing and debugging. Several respondents noted that frequent updates to SDKs, hardware APIs, and simulation environments introduce regressions or behavioral inconsistencies, a pattern strongly reflected in Section 7.2 and Section 7.3, where over 80% of the participants reported experiencing bugs at least occasionally due to platform updates. Interview Respondent 4 observed that "[W]hen we're developing something that's truly new, the standard tools aren't necessarily there ... but automated testing is probably [used] once things are working or near working, and you just want to make sure it hasn't broken." This reflects a workflow where developers are reactive—writing tests to prevent regressions—rather than proactively guided by robust test frameworks or continuous integration infrastructure.

*Mitigation strategies.* To address instability in libraries and platforms, developers emphasized the importance of version control practices such as pinning dependencies, maintaining local mirrors, and delaying upgrades until stability is confirmed. Some reported building internal safeguards by testing updates in controlled environments before integrating them into production workflows.

Tool support could go further by providing automated compatibility checking, migration assistants, and clearer deprecation policies within SDKs. Such practices would help reduce the regressions and inconsistencies that more than 80% of our respondents reported experiencing.

*8.3.3 Non-Determinism and Scalability.* Debugging is further complicated by the probabilistic nature of QC. Unlike classical programs, quantum programs often fail silently, producing output that appears valid, but is statistically incorrect. Developers must resort to repeated execution and statistical inference to determine whether a bug exists. As Interview Respondent 4 explained, “There are things that you take for granted with classical software... You can’t do [some things] in a quantum algorithm except under very specialized circumstances.” In this context, simulators play an important role: several respondents explicitly reported relying on simulation tools to support debugging and validation activities (see Section 6.1). However, classical simulators are fundamentally constrained by the exponential scaling of quantum state spaces. As circuits grow in qubit count or depth and as noise models are introduced, simulation becomes prohibitively slow or even infeasible. Several participants noted that simulation-based debugging no longer scales well with modern applications, especially when trying to model realistic execution or perform statistical validation. These constraints leave developers without reliable, scalable alternatives for evaluating the correctness on actual hardware.

*Mitigation strategies.* To cope with non-determinism, many respondents reported relying on repeated executions combined with statistical inference, examining whether output distributions remained stable across runs. Some also used simulators as reference models for smaller circuits, treating deviations between simulated and hardware behavior as potential indicators of noise or defects. In response to scalability limits, developers emphasized modular validation, testing smaller subcircuits in isolation before integration, and in most cases relying on visualization techniques as part of their testing practices for larger circuits. Although uneven in adoption, these practices highlight the resourcefulness of practitioners in the face of tool limitations. Looking ahead, the literature suggests that approaches such as metamorphic testing [60], mutation-based frameworks [24, 52], and noise-aware simulation techniques [84] may offer more systematic support for managing probabilistic behavior and scaling challenges.

*8.3.4 Cross-Layer Complexity.* The hybrid nature of many quantum applications introduces additional barriers. Developers must orchestrate quantum and classical components across multiple layers, with bugs often emerging at the interface between them. As discussed in Section 6, debugging across these layers typically requires trial-and-error reasoning, particularly in the presence of quantum noise or orchestration logic failures. Without unified tooling to trace errors across domains or diagnose probabilistic behavior, even identifying whether a bug exists can become a nontrivial task.

*Mitigation strategies.* Cross-layer complexity in hybrid applications may be alleviated through unified debugging environments that can trace both classical orchestration and quantum execution within the same workflow. Respondents noted the difficulty of diagnosing whether failures originated in classical pre/post-processing or within the quantum hardware itself. Promising approaches include integrated logging and tracing across layers, runtime monitors that separate classical from quantum errors, and CI/CD pipelines tailored to hybrid systems. These kinds of cross-cutting capabilities were notably absent in current practice, but they could reduce reliance on *ad hoc* trial-and-error reasoning.

*8.3.5 Implications.* Collectively, these barriers suggest that current quantum development workflows place significant cognitive and organizational burden on developers. Although traditional testing practices such as unit and regression testing are commonly used, they are insufficient to address the unique challenges of quantum software. Without improvements in abstraction, tool integration, runtime observability, and hardware-level diagnostics, developers will remain reliant on manual, error-prone processes that struggle to scale with growing application complexity.

### 8.4 Tooling Gaps, Developer Desiderata, and Research Implications

Quantum developers today are not short on effort—they are short on tools that meet the practical demands of hybrid quantum-classical development. Across testing and debugging, our survey revealed persistent friction, unmet needs, and a strong appetite for tools that integrate more cleanly with real-world workflows. These findings offer important insights for tool builders and researchers alike.

**Testing and Debugging Support.** As noted in Section 6.1, only one respondent reported using a dedicated quantum debugging tool. Most relied on classical techniques such as print statements, manual inspection, and trial-and-error debugging. Even the utilities provided by leading frameworks such as Qiskit's Test Utilities were underutilized, and many developers were unaware of the academic offerings. This reflects a disconnect not necessarily in technical merit, but in usability, integration, and awareness.

The respondents seemingly called for tools that are more intuitive, more scalable, and better aligned with the unique challenges of quantum software. Particularly valuable would be debugging and testing frameworks that:

- Automatically validate output against known expectations or templates.
- Surface statistical anomalies in repeated circuit executions.
- Localize errors within composite or deeply nested quantum programs.
- Differentiate between classical and quantum sources of failure.

These capabilities are essential given the probabilistic and hybrid nature of quantum systems and the echo challenges documented in previous work [24, 83].

**Visualization and Simulation Aids.** Circuit visualizations are widely used but scale poorly as quantum programs grow in size and complexity. Several participants noted that current tools are best suited for educational settings or small-scale debugging. Developers appeared to need:

- Hierarchical or interactive visualizations that reveal structure while hiding low-level details.
- Visualization of output distributions, entanglement, or intermediate state behavior.
- Better integration of simulation with testing and introspection tools.

Together, these capabilities could mitigate limited observability in quantum systems and better support developers in reasoning about program behavior across multiple abstraction layers as quantum circuits and applications scale.

**Versioning, Dependency Management, and Stability.** Library and platform evolution emerged as a major source of instability. Over 80% of developers reported encountering bugs at least occasionally after library updates, with 42% experiencing them frequently or very frequently. These issues were most often tied to broken dependencies, undocumented changes, or incompatible API shifts—particularly in frameworks like Qiskit [14, 90]. Developers emphasized the need for:

- Dependency managers that resolve compatibility issues.
- Version pinning, rollback, and migration tooling.
- Automatic detection of breaking changes with contextual guidance.
- Consistent APIs and clear deprecation policies with sufficient notice.
- Clear documentation that is regularly maintained.

These problems have led many developers to delay updates or treat them as high-risk operations, increasing maintenance overhead and slowing development workflows. Collectively, the needs articulated by respondents point to versioning and dependency management as critical pain points where targeted tooling and clearer platform practices could substantially reduce friction in QSE.

**Workflow Integration and Ecosystem Support.** A recurring theme in our findings was the fragmented nature of the quantum development ecosystem. Many developers reported building in-house tools to fill gaps, noting poor integration across the development stack. Respondents appeared to express a desire for:

- Unified environments that simplify coordination between classical and quantum workflows.
- IDE integrations for debugging, visualization, and test management.
- Logging, tracing, and introspection tools that span both execution layers.

Such fragmentation has led many developers to rely on bespoke or in-house solutions to bridge gaps across the development stack, increasing cognitive overhead and slowing debugging cycles. Collectively, respondents' needs point to workflow integration and ecosystem support as critical areas where better coordination between classical and quantum tooling could substantially improve developer productivity and reduce complexity.

**Implications for Tool Builders and Researchers.** Tool researchers face a dual challenge: addressing the unique demands of quantum development while also aligning with the practices and constraints of real-world developers. Many promising academic tools suffer from poor adoption not because of weak technical foundations, but due to steep learning curves, fragile integrations, or lack of validation in production-like settings.

Our results underscore the importance of participatory design and practical validation. Developers are more likely to adopt tools that are:

- Easy to learn and integrate.
- Stable across framework updates.
- Tailored to hybrid architectures.
- Maintained with strong documentation and examples.

Echoing recent calls in the literature [19, 37], we argue that future efforts in QSE tooling should prioritize not only innovation but developer adoption. Bridging the gap between research and practice will require close collaboration across quantum software researchers, platform maintainers, and industry practitioners.

Taken together, these recommendations suggest different priorities for different stakeholders. Tool builders and platform maintainers can have the greatest near-term impact by improving integration with mainstream SDKs, ensuring stability across updates, and offering layered abstractions and IDE support that reduce cognitive overhead. Researchers should focus on advancing promising approaches from the literature—such as metamorphic and mutation testing [24, 52, 60]—while validating them in production-like environments and co-designing with practitioners to improve adoption. Practitioners themselves can accelerate progress by adopting practices such as version pinning, regression testing, and modular validation, while providing feedback to tool builders on usability and workflow fit. By clarifying these roles and responsibilities, our study highlights a roadmap where immediate impact lies with integration and stability improvements by tool builders, medium-term progress depends on research innovation aligned with practice, and long-term change will be sustained by practitioner uptake and community-driven best practices.

### 8.5 Educational Backgrounds and Engineering Gaps

QSE is fundamentally interdisciplinary, demanding fluency in both quantum mechanics and classical software engineering. As described in Section 4.4, our participant pool included developers with diverse academic backgrounds and experience levels, most commonly in physics, computer science, and engineering. This range of educational training reflects the field's dual roots in quantum science and traditional computing, and suggests that effective quantum software education must bridge theoretical and practical domains.

Despite high academic achievement where 18 out of 26 respondents held a PhD (Figure 4), only nine respondents reported receiving formal training in quantum software engineering (Figure 6). This lack of structured, cross-domain education creates a practical skills gap: many developers possess deep theoretical knowledge or strong programming skills, but not both. As Interview Respondent 3 emphasized: "It's very hard to find someone who both has the knowledge of quantum physics, even quantum information, and is a competent software engineer." This challenge underscores a core tension in the field: bridging the gap between scientific foundations and engineering practice.

This divide contributes to inconsistencies in development practices. Participants with physics backgrounds often rely on mathematical models and manual inspection, while those with software engineering backgrounds apply classical tooling workflows, sometimes without sufficient awareness of quantum-specific phenomena. These different perspectives help explain the heterogeneous testing and debugging practices observed in Sections5 and Section 6, including the dependence on visualization, trial and error methods, and informal validation strategies.

Our findings point to an urgent need for cross-disciplinary education and inclusive tool design. Educational programs must equip physicists with applied software engineering skills and introduce computer scientists to quantum reasoning. Tool builders, meanwhile, should accommodate a range of user backgrounds by offering layered abstractions, contextual error guidance, and built-in documentation. Bridging these educational and practical divides is essential for scaling the quantum workforce and building reliable software systems. As noted in prior work [19], tool usability and developer productivity depend not just on technical capabilities, but also on the alignment with the mental models and needs of different users.

### 8.6 Roadmap for Tool Adoption and Best Practices

Our findings indicate that while developers are heavily involved in testing and debugging, the adoption of quantum-specific tools designed for these tasks remains limited. Based on these insights, we propose a roadmap for fostering tool adoption and establishing best QSE practices. This roadmap outlines practical steps across three complementary dimensions: awareness, integration, and usability.

*8.6.1 Awareness.* Many participants were unaware of existing academic or SDK-provided tools, underscoring the need for greater visibility. Outreach through documentation, tutorials, and community-driven examples can help bridge this gap. Tool builders should prioritize clear, accessible onboarding materials that target developers from both physics and software backgrounds.

*8.6.2 Integration.* Developers consistently emphasized that the tools must fit naturally into their workflows. Best practices here include embedding testing and debugging capabilities directly into widely used SDKs (*e.g.*, Qiskit, Cirq, Pennylane, *etc.*) and providing lightweight APIs for integration into CI/CD pipelines. Tools that seamlessly connect classical and quantum components are more likely to be adopted in production settings.

*8.6.3 Usability.* Respondents frequently cited poor scalability, limited flexibility, and unclear interfaces as barriers to adoption. Frequent SDK and platform updates also introduced regressions and compatibility issues, which created instability for developers. To address these concerns, tool design should prioritize:

- Hierarchical visualizations that scale with system size.
- Statistical testing utilities that automate distribution analysis.
- Built-in dependency and versioning checks to mitigate update-related regressions.
- Documentation and examples that reflect real-world development scenarios.

*8.6.4 Best Practices.* Until specialized tools mature, practitioners can benefit from a layered approach to testing and debugging:

(1) Start with simulation and unit testing for early correctness checks.
(2) Apply regression testing across updates to catch breakages in SDKs and APIs.
(3) Incorporate statistical validation (*e.g.*, repeated runs, fidelity checks) to account for probabilistic outcomes.
(4) Adopt version pinning and dependency management as standard practice to reduce instability from frequent platform changes.

Together, these steps provide a pragmatic roadmap: Increase awareness of available tools, integrate them tightly into mainstream workflows, and improve usability through participatory design. Establishing shared best practices around layered testing, statistical validation, and stable dependency management can accelerate tool adoption and contribute to a more mature and sustainable QSE ecosystem.

### 8.7 Future Directions

Our findings highlight a clear set of priorities for advancing QSE: closing the gap between academic tools and real-world usage, increasing the abstraction and ease of use of quantum libraries and tools, addressing hybrid workflow needs, and strengthening cross-disciplinary education. Developers in our study were highly engaged in testing and debugging, yet tool adoption remained fragmented and limited. While traditional practices such as unit testing, manual inspection (*e.g.*, logging), and simulation were prevalent, quantum-specific tools developed in academia—such as Muskit [52], QmutPy [24], and QuCAT [85]—were rarely used. As one respondent (Interview Respondent 3) explained, "We have our own internal [tools]... academic tools seem interesting, but they don't fit our workflows." To increase impact, researchers must explore why developers default to custom solutions and prioritize co-design, usability testing, and integration with common SDKs like Qiskit, Cirq, and PennyLane.

Looking ahead, more robust development and support is needed for testing and debugging in quantum software. As discussed in Section 8.3, simulators play an important role in current development workflows, with many survey participants reporting reliance on simulation tools to support testing and debugging activities. However, respondents also noted the limits of classical simulators—particularly as circuits grow deeper and noise modeling becomes essential. Developers described simulators becoming "increasingly slow or unusable," underscoring the need for runtime-aware diagnostics and execution logging that can operate directly on hardware. Similarly, given the prevalence of probabilistic failure modes, tools should offer statistical testing capabilities that go beyond manual histogram inspection and domain intuition.

Another area of future potential research is to examine testing and debugging practices by application domain. While our survey did not explicitly categorize quantum software by type (*e.g.*, machine learning, chemistry, optimization, *etc.*), open-text responses suggested that developers are already applying domain-specific reasoning in their testing workflows. As quantum computing matures and begins to see broader deployment across application areas, it will be important to investigate whether certain domains adopt distinctive testing strategies, face unique challenges, or require specialized tools. Understanding these differences could inform the design of domain-aware methodologies and strengthen the generalizability of best practices in QSE.

A related and recurring theme across our findings is the need for improved abstraction in quantum software development. Many of the challenges reported by respondents—such as manual testing, difficulty reasoning about circuit behavior, and developer errors rooted in low-level misunderstandings—stem from the need to operate at very low levels of abstraction that quantum libraries and platforms currently provide. While low-level access is essential for research and experimentation, practitioners emphasized the lack of higher-level constructs that encapsulate common patterns, manage complexity, and reduce the cognitive burden of reasoning about quantum behavior directly. Advancing abstraction in quantum libraries and tooling—while preserving transparency and control—should therefore be a dedicated focus of future research, with the goal of enabling developers to reason at higher levels of intent rather than low-level implementation details, and of reducing error rates and reliance on *ad hoc* workflows.

Finally, future work should address the skill gap between quantum theory and software engineering. Although most participants held PhDs, few had received formal training in quantum software engineering, contributing to inconsistent practices and cognitive overhead. Tools that provide layered abstractions and onboarding support, combined with educational programs that bridge the physics-CS divide, will be essential for scaling the workforce. By aligning tools with developer needs, integrating them into daily workflows, and fostering collaboration between academia and industry, the QSE community can build a foundation for more reliable, maintainable, and accessible quantum software.

## 9 THREATS TO VALIDITY

This study may contain several potential threats to validity.

### 9.1 Internal Validity

One limitation of our study is self-reporting bias in the survey and interview responses. Developers may overestimate or underestimate their experience, leading to potential inaccuracies in our data set. Additionally, selection bias could be present, since the sample primarily consists of developers who are already involved with quantum software engineering communities. This may lead to an overrepresentation of certain perspectives while omitting others, such as individuals who have struggled with quantum development and disengaged from the field.

### 9.2 External Validity

The generalizability of our findings is limited by our sample size and demographic composition. Although we collected responses from a diverse set of quantum developers, the survey was predominantly completed by individuals with backgrounds in physics, mathematics, and computer science. As a result, our findings may not fully capture the perspectives of industry practitioners or software engineers transitioning from classical to quantum computing. Furthermore, our study focuses on current quantum programming frameworks and hardware, so our conclusions may not generalize to future developments such as fault-tolerant quantum computing, new programming paradigms, or significant shifts in toolchain architecture.

Moreover, like many empirical studies in emerging domains, our findings are based on a modest sample size—26 survey respondents and four follow-up interviews. This sample provides valuable first-hand insights but also imposes limits on generalizability. In particular, the sample may over-represent highly engaged practitioners (*e.g.*, those active in research or industry communities) while under-representing developers who work in less visible or proprietary settings. Percentages should therefore be interpreted as indicative of emerging patterns rather than definitive measurements of prevalence. For example, when eight of 26 respondents reported using metamorphic testing, this suggests an emerging practice but does not establish prevalence across the community.

That said, many of the themes we identified—such as reliance on unit testing, limited adoption of specialized debugging tools, the fragility of SDKs, and the difficulty of handling non-determinism—echo findings in prior work [7, 37, 46, 59, 64]. This triangulation increases our confidence that the challenges we describe are not artifacts of our sample alone, but recurring issues across contexts. We therefore view our study as an exploratory contribution: it documents recurring problems and coping strategies, surfaces gaps between research tools and practice, and motivates future large-scale replications.

### 9.3 Construct Validity

We took several steps to ensure that our study design accurately captured the constructs we were aiming to measure. Guided by best practices in empirical software engineering, we carefully formulated our survey and interview questions to reduce bias, avoid leading language, and align clearly with our research objectives. The instruments were iteratively refined through pilot testing and internal review to ensure clarity and construct alignment.

To further support construct validity, we triangulated our data sources—combining quantitative survey results with qualitative interview insights—to surface robust patterns and minimize overreliance on any single method. We also used a multicoder strategy in analyzing interview responses: multiple researchers independently coded transcripts, followed by iterative rounds of discussion to refine the codebook and reconcile disagreements. This approach helped to ensure that our thematic analysis reflected the intended meanings of the participants rather than the assumptions of the individual researcher.

### 9.4 Conclusion Validity

We designed our analysis to minimize bias and improve the credibility of our conclusions. By triangulating survey and interview data and employing collaborative coding practices, our goal was to ensure that the patterns we identified were representative and well supported. Although we do not claim causal relationships between specific variables, the convergence of evidence between data sources and participants gives us confidence in the validity of our reported themes.

Although our data was collected beginning in mid-2024, we believe that our conclusions remain timely and relevant. The core challenges identified, such as limited tool adoption, manual debugging, versioning difficulties, and integration obstacles, continue to appear in more recent studies [9, 19, 37, 57, 81]. This persistence suggests that these are structural issues within the quantum software ecosystem, not artifacts of a specific moment in time.

## 10 CONCLUSIONS

Quantum software engineering is marked by distinct challenges, ranging from probabilistic behavior and hardware limitations to immature tooling and low-level abstractions. Our study found that while developers are highly engaged in testing and debugging, their workflows are constrained by limited tool support and a heavy reliance on classical practices like print statements and visual inspection. The scarcity of quantum-specific debugging tools and scalable testing frameworks forces practitioners to adopt manual and often error-prone approaches.

Our study also found recurring issues with platform / library instability, particularly in response to frequent updates and changing APIs. Developers reported that changes in widely used frameworks like Qiskit often introduce regressions, break dependencies, and lack clear migration paths. These pain points, combined with the need to debug in many cases across quantum and classical subsystems, amplify the difficulty of identifying, isolating, and resolving bugs in quantum software.

To advance the state of QSE, future efforts must prioritize usability, integration, and developer-centered design. This includes building tools that scale with quantum program complexity, support statistical validation, and provide better insight into circuit behavior. By aligning research efforts with real-world developer needs, the community can close the gap between theory and practice and lay the foundation for reliable and maintainable quantum software development.